\documentclass[aps,prx,twocolumn,superscriptaddress]{revtex4-2}
\usepackage[english]{babel} %%% 'french', 'german', 'spanish', 'danish', etc.
\usepackage{amssymb}
\usepackage{amsmath}
\usepackage{mathtools}
\usepackage{txfonts}
\usepackage{mathdots}
\usepackage[normalem]{ulem}
\usepackage[dvips]{graphicx}
\usepackage{epsfig}
\usepackage{graphicx}
\usepackage{array}
\usepackage{amssymb}
\usepackage{amsfonts}
\usepackage{amsmath}
\usepackage{mathrsfs}
\usepackage{booktabs}
\usepackage{threeparttable}
\usepackage{multirow}
\usepackage{subfigure}
\usepackage{epsfig}
\usepackage{threeparttable}
\usepackage{chngpage}
\usepackage{float}
\usepackage{xcolor}
\usepackage{bm}
\usepackage{graphicx}
\usepackage[toc]{appendix}
\usepackage{tabularx}
\usepackage{graphicx}
\usepackage{adjustbox}
\usepackage{comment}
\usepackage{hyperref}
\hypersetup{
    unicode=false,     % non-Latin characters in Acrobat bookmarks
    pdftoolbar=false,  % show Acrobat toolbar?
    pdfmenubar=true,   % show Acrobatmenu?
    pdffitwindow=false, % window fit to page when opened
    pdfstartview={FitH},% fits the width of the page to the window
    pdftitle={},    % title
    pdfauthor={Authors},     % author
    pdfsubject={},   % subject of the document
    pdfcreator={},   % creator of the document
    pdfproducer={}, % producer of the document
    pdfkeywords={quantum many-body scars} {superconducting processor} {quantum state tomography}, % list of keywords
    pdfnewwindow=true,% links in new window
    colorlinks=true,% false: boxed links; true: colored links
    linkcolor=black,% color of internal links (change box color with linkbordercolor)
    citecolor=blue, % color of links to bibliography
    filecolor=magenta,% color of file links
    urlcolor=blue% color of external links
}
\newcommand{\beginsupplement}{%
    \setcounter{table}{0}
    \renewcommand{\thetable}{S\arabic{table}}%
    \setcounter{figure}{0}
    \renewcommand{\thefigure}{S\arabic{figure}}%
    \setcounter{equation}{0}
    \renewcommand{\theequation}{S\arabic{equation}}%
    \setcounter{section}{0}
    \renewcommand{\thesection}{\arabic{section}}%
   }

\begin{document}

\title{Observation of many-body Fock space dynamics in two dimensions}
\author{Yunyan Yao}
\thanks{These authors contributed equally}
\author{Liang Xiang}
\thanks{These authors contributed equally}
\author{Zexian Guo}
\author{Zehang Bao}
\affiliation{ZJU-Hangzhou Global Scientific and Technological Innovation Center, Department of Physics, Interdisciplinary Center for Quantum Information, and Zhejiang Province Key Laboratory of Quantum Technology and Device, Zhejiang University, Hangzhou, China}
\author{Yong-Feng Yang}
\affiliation{School of Physical Science and Technology, Lanzhou Center for Theoretical Physics, and Key Laboratory of Theoretical Physics of Gansu Province, Lanzhou University, Lanzhou, Gansu 730000, China}
\author{Zixuan Song}
\author{Haohai Shi}
\author{Xuhao Zhu}
\author{Feitong Jin}
\author{Jiachen Chen}
\author{Shibo Xu}
\author{Zitian Zhu}
\author{Fanhao Shen}
\author{Ning Wang}
\author{Chuanyu Zhang}
\author{Yaozu Wu}
\author{Yiren Zou}
\author{Pengfei Zhang}
\author{Hekang Li}
\author{Zhen Wang}
\author{Chao Song}
\affiliation{ZJU-Hangzhou Global Scientific and Technological Innovation Center, Department of Physics, Interdisciplinary Center for Quantum Information, and Zhejiang Province Key Laboratory of Quantum Technology and Device, Zhejiang University, Hangzhou, China}

\author{Chen Cheng}
\affiliation{School of Physical Science and Technology, Lanzhou Center for Theoretical Physics, and Key Laboratory of Theoretical Physics of Gansu Province, Lanzhou University, Lanzhou, Gansu 730000, China}
\author{Rubem Mondaini} 
\affiliation{Beijing Computational Science Research Center, Beijing 100094, China}

\author{H. Wang}
\author{J. Q. You}
\author{Shi-Yao Zhu} 
\author{Lei Ying}
\email{leiying@zju.edu.cn}
\author{Qiujiang Guo}
\email{qguo@zju.edu.cn}
\affiliation{ZJU-Hangzhou Global Scientific and Technological Innovation Center, Department of Physics, Interdisciplinary Center for Quantum Information, and Zhejiang Province Key Laboratory of Quantum Technology and Device, Zhejiang University, Hangzhou, China}

\date{\today}

\begin{abstract}
Quantum many-body simulation provides a straightforward way to understand fundamental physics and connect with quantum information applications. However, suffering from exponentially growing Hilbert space size, characterization in terms of few-body probes in real space is often insufficient to tackle challenging problems such as quantum critical behavior and many-body localization (MBL) in higher dimensions. Here, we experimentally employ a new paradigm on a superconducting quantum processor, exploring such elusive questions from a Fock space view: mapping the many-body system onto an unconventional Anderson model on a complex Fock space network of many-body states. By observing the wave packet propagating in Fock space and the emergence of a statistical ergodic ensemble, we reveal a fresh picture for characterizing representative many-body dynamics: thermalization, localization, and scarring. In addition, we observe a quantum critical regime of anomalously enhanced wave packet width and deduce a critical point from the maximum wave packet fluctuations, which lend support for the two-dimensional MBL transition in finite-sized systems. Our work unveils a new perspective of exploring many-body physics in Fock space,  demonstrating its practical applications on contentious MBL aspects such as criticality and dimensionality. Moreover, the entire protocol is universal and scalable, paving the way to finally solve a broader range of controversial many-body problems on future larger quantum devices.
\end{abstract}

\maketitle

\section*{Introduction}

\begin{figure*}
\includegraphics[width=1.8\columnwidth]{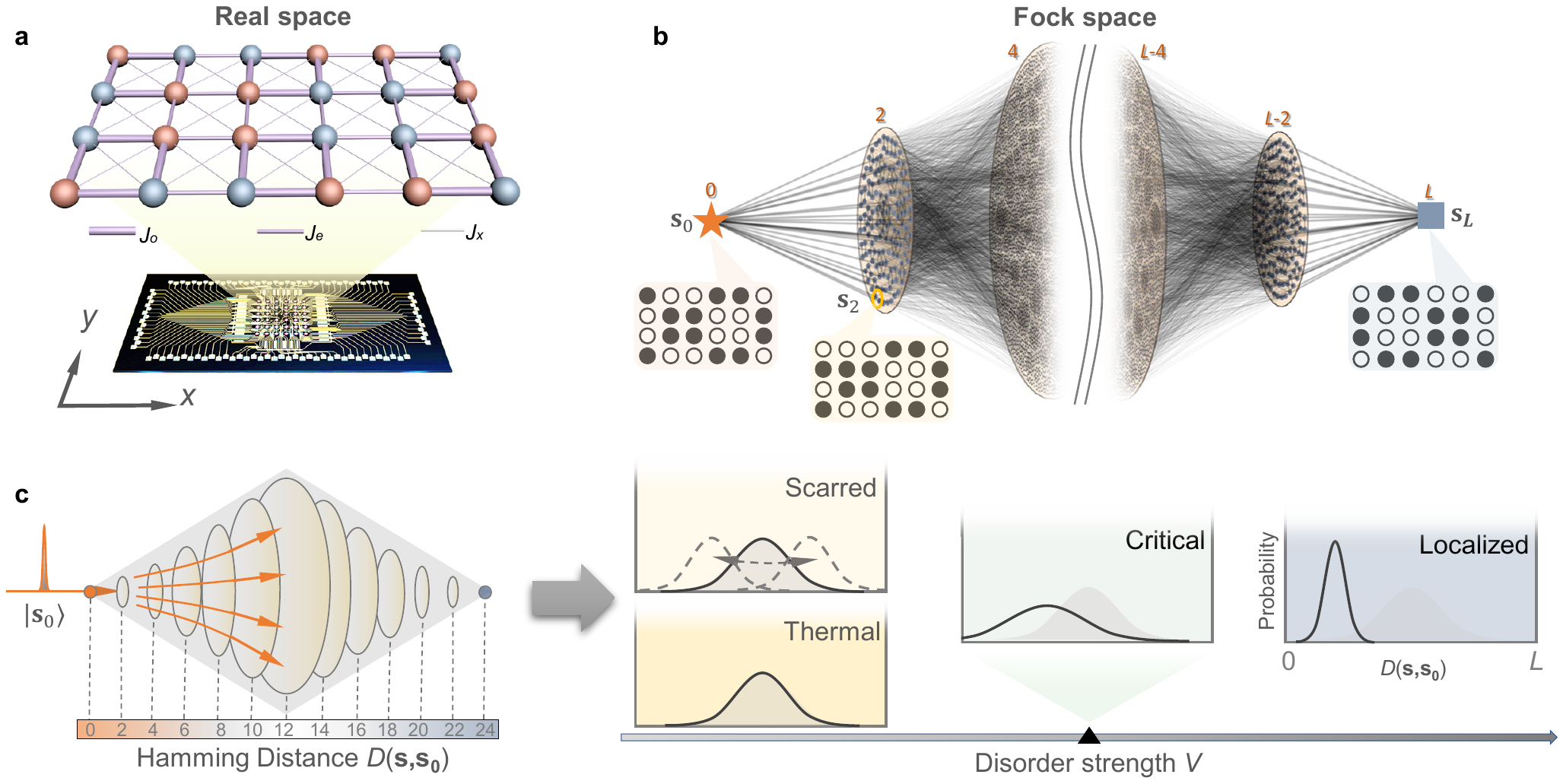}
    \caption{
    {\bf Quantum processor and schematic of many-body dynamics in Fock space.} 
    {\bf a}, 
    The lower panel shows the optical micrograph of a $36$-qubit superconducting quantum processor. Using $24$ of them, we emulate a hard-core Bose-Hubbard model on a $4\times6$ two-dimensional lattice~(upper panel). Each red (blue) sphere stands for a real-space qubit at the excited (ground) state, and lines represent the photon hopping strength connecting two sites, whose thickness indicates the programmable coupling strength. 
    {\bf b}, 
    Fock-space representation of the real-space lattice in {\bf a}, where each site $\mathbf{s}$ (denoted by a dot) represents a specific photon excitation configuration in the real-space lattice. 
    The extensive local connectivity (gray lines) signifies all possible one-photon hoppings in Hamiltonian~(\ref{eq:hamiltonian}). 
    The initial configuration $\mathbf{s}_0$ is the apex of the network, marked by an orange star. All other sites are arranged according to the Hamming distance $D(\mathbf{s}, \mathbf{s}_0)$ away from it. The yellow circle ($\mathbf{s}_2$) and the blue square~($\mathbf{s}_L$) are two representative configurations corresponding to $D(\mathbf{s}, \mathbf{s}_0)=2$ and $L$, respectively. 
    {\bf c}, Fock-space visualization of the typical dynamics.
    A far-from-equilibrium initial state $|\mathbf{s}_0\rangle$ is prepared as a Fock state, whose wave packet is a unit impulse function located at $D=0$. The unitary dynamics result in its propagation and diffusion in the Fock-space network over time, whose long-time behavior describes the equilibration properties. At weak disorder regime, the system shows a thermalizing behavior, and the wave packet quickly reaches the ergodic distribution (shaded gray region) with a maximum occurring at $L/2$ for the majority of states. Few scarred states, however, exhibit pendulum-like dynamics with slow thermalization as visualized by the dashed curves and arrows. At strong disorders, the propagation of the wave packet is suppressed at $D(\mathbf{s}, \mathbf{s}_0)<L/2$, and its width is narrower than the ergodic case. Around the critical point (black triangle), the wave packet exhibits a broader width, implying stronger fluctuation.   
    } 
    \label{fig:1}
\end{figure*}

Strong-correlated particles in an isolated quantum system with appropriate perturbations trigger abundant physical phenomena, typified by the many-body version of quantum thermalization~\cite{Deutsch1991PRA, Srednicki1994PRE,Rigol2008Nature,DAlessio2016AP}, localization~\cite{Anderson1958PR, Basko2006AP,Rahul2015annu,Abanin2019RMP,ALET2018CRP}, scarring~\cite{Turner2018np,Serbyn2021np}, etc. Their experimental realizations in various platforms~\cite{Schreiber2015science,Smith2016np, Kai2018PRL, Roushan2017science} arouse great interest ranging from fundamental physics to  applications in quantum information, in particular for being examples of potential quantum advantages in many-body quantum simulations~\cite{Xiao2021science}. Although tremendous experimental efforts have been devoted to understanding quantum thermalization~\cite{Neill2016np, Kaufman2016science} and its breakdown~\cite{Schreiber2015science,Smith2016np, Lukin2019science}, there is still no strong consensus on further open problems such as the stability of many-body localization (MBL) phase in higher dimensions~\cite{Roeck2017PRB,Potirniche2019PRB,Doggen2020PRL} and critical properties of the MBL transition~\cite{Potter2015PRX,Khemani2017PRX,Dumitrescu2019PRB}. So far, existing experimental explorations on these topics have been largely confined to the conventional framework in real space, focusing on the characterization of the dynamics of few-body observables such as the imbalance and correlations~\cite{Choi2016science,BordiaPRX2017,Rispoli2019science}, rather than the bipartite entanglement entropy (EE). The latter, a global probe often used in theoretical studies since it distinctively describes the many-body properties encoded in the quantum states, is, on the other hand, extremely challenging for experiments. Thus,  strong motivation exists to look for new experimental playgrounds to resolve these elusive problems.

On a different perspective, the many-body problem can be recast as a virtual ``particle'' hopping on a complex Fock-space network~\cite{Basko2006AP, Welsh2018JoConMatt} with extensive local connectivity, each site of which represents a many-body state. This idea originated from mapping the complex disordered quantum dot system to an Anderson localization problem in Fock space associated with many-body states~\cite{Altshuler1997PRL}, where the nonergodic behavior is represented by the localization in Fock space under strong-correlated disorders~\cite{Basko2006AP,Nicolas2019PRL,Logan2019,Tomasi2019PRB}. This new angle has led to a series of novel insights, such as multi-fractal scaling of MBL eigenstates~\cite{Nicolas2019PRL}, emergent Hilbert space fragmentation~\cite{Tomasi2019PRB}, and nonergodic extended phase~\cite{Luca2014PRL, Tomasi2019spp,Wang2021PRL}. Nevertheless, experimental investigations along this line are still scarce, since the conventional analysis in Fock space largely relies on the quantification of the experimentally prohibitive inverse participation ratios (IPR)~\cite{Nicolas2019PRL}.

To shed light on these long-standing challenges, here, we experimentally demonstrate a new paradigm of probing many-body dynamics in Fock space, describing a universal and scalable protocol capable of exploring such controversial problems, as MBL in higher dimensions and associated quantum criticality. The underlying idea is illustrated in Fig.~\ref{fig:1}. An isolated many-body system in real space (the upper panel of Fig.~\ref{fig:1}{\bf a}) can be equivalently represented in Fock space (Fig.~\ref{fig:1}{\bf b}), where each Fock-space site corresponds to a photon excitation configuration $\mathbf{s}$ in real space and their connectivities denote all allowed hoppings. Thus, the many-body dynamics of a far-from-equilibrium initial Fock (product) state $|\mathbf{s}_0\rangle$ can be regarded as a wave packet moving along the radial direction of the Fock-space network labeled by the Hamming distance $D(\mathbf{s},\mathbf{s}_0)=\sum_{m,n}|s^{mn}-s_{0}^{mn}|$ (the left panel of Fig.~\ref{fig:1}{\bf c}), where $s^{mn}=0$ or $1$ denotes the real-space site $(m,n)$ being either in the ground or excited state. 

Within this Fock-space approach, we experimentally characterize typical many-body dynamics and further explore contentious areas of the MBL transition on a two-dimensional (2D) superconducting qubit array. We track how the wave packet propagates on the Fock-space network, providing an intuitive physical picture of thermalization and its breakdown (localization and scarring) from the Fock-space view (right panel of Fig.~\ref{fig:1}{\bf c}). Quantum ergodicity is further qualified by the Bhattacharyya metric, and we find that only a finite fraction of the Fock space is actively involved at large disorder strengths, a hallmark of Fock-space localization. Moreover, the anomalous nonmonotonic behavior of wave packet width with growing disorder allows us to quantitatively identify a three-regime phase diagram for the finite-size 2D nonergodic transition, which is hard to capture for conventional real-space experimental observations. The critical disorder $V_c$ is extracted by the maximum of wave packet fluctuations in the disorder dependence. Remarkably, it agrees well with the numerical value by means of the experimentally inaccessible probe, the bipartite EE, further confirming the effectiveness of our protocol.

\begin{figure*}
    \includegraphics[width=2\columnwidth]{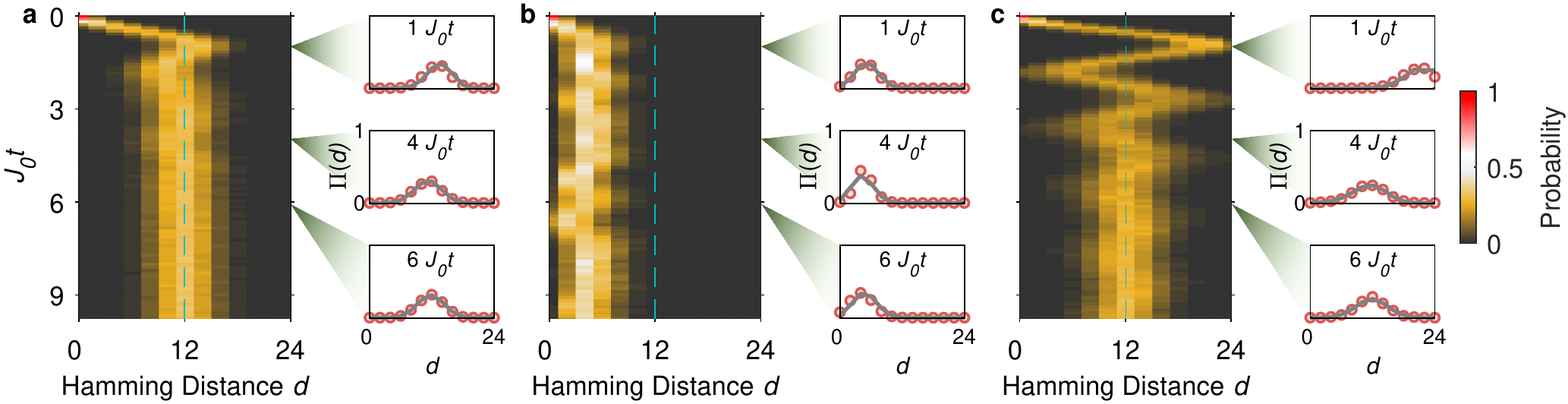}
    \caption{{\bf Many-body dynamics in Fock space.} 
    Plotted are experimental \emph{dynamical} radial probability distribution $\Pi(d, t)$ in Fock space, with color bar on the far right, which characterize dynamics of  thermalization ({\bf a}), localization ({\bf b}), and scarring ({\bf c}). Insets are the snapshots showing the comparison of Fock-space wave packets for experiments (dots) and numerics (curves).
    }
    \label{fig:2}
\end{figure*}

\section*{Experimental platform and protocol}
To reveal many-body dynamics in Fock space with the aforementioned approach, we utilize a 2D flip-chip superconducting quantum processor (the lower panel in Fig.~\ref{fig:1}{\bf a}). It provides a 6$\times$6 transmon qubit lattice with long qubit energy relaxation times ($T_1\sim 120~\mu $s), high-fidelity site-resolved controllability (single-qubit fidelity $\sim$ 0.997), and tunable interactions (Supplementary Sections 1, 2). Using a subset of $L=24$ qubits, we effectively emulate a 4$\times$6 hard-core Bose-Hubbard lattice in real space (the upper panel of Fig.~\ref{fig:1}{\bf a}) and its Hamiltonian (Supplementary Section 4) is given by
\begin{equation}\label{eq:hamiltonian} 
\begin{split}
    \frac{H_\mathrm{R}}{\hbar} = & \sum_{  m,n } \left( \chi_{m} \sigma^+_{m,n} \sigma^-_{m+1,n}    + \gamma_{n} \sigma^+_{m,n} \sigma^-_{m,n+1}   +   \mathrm{h.c.}  \right)   \\
    &+  \sum_{m,n}  V_{m,n} \sigma^+_{m,n} \sigma^-_{m,n} + {H}_\mathrm{x}
\end{split}
\end{equation}
where $\sigma_{m,n}^{+}$ ($\sigma_{m,n}^{-}$) is the two-level raising (lowering) operators of qubit $(m,n)$ and $\hbar$ is the reduced Planck constant. The first term in Eq.~(\ref{eq:hamiltonian}) describes the kinetic motion of bosons, where $\chi_{m}$ and $\gamma_{n}$ denote the strength of tunable nearest-neighbor hoppings along  $x$- and $y$-axis respectively. The second term represents the local bosonic occupation energy, which can be adjusted individually for each site $(m,n)$ by tuning the qubit frequency. To mimic a 2D disordered system, $V_{m,n}$ is chosen from the uniform random distribution $[-V, V]$. A perturbation term, $H_\mathrm{x}=g_\mathrm{x}\sum_{m,n}\left(\sigma^+_{m,n}\sigma^-_{m+1,n+1} + \sigma^+_{m,n}\sigma^-_{m+1,n-1}  +\mathrm{h.c.}\right)$, is associated to parasitic cross hoppings that naturally occur in the device. As visualized in the upper panel of Fig.~\ref{fig:1}{\bf a}, we set $J_\mathrm{o}/2\pi=2J_\mathrm{e}/2\pi\approx-6$~MHz and  $g_\mathrm{x}/2\pi\approx0.9$~MHz in the model to realize a 2D Su-Schrieffer-Heeger (SSH) model~\cite{Benalcazar2017science} (Supplementary Sections 4, 5). In the half-filling condition, this setup naturally makes the Hamiltonian (\ref{eq:hamiltonian}) a non-integrable model rather than the non-interacting single-particle problem~\cite{Karamlou2022npj}. 

From the Fock-space view, the $4\times6$ real-space 2D Hamiltonian (\ref{eq:hamiltonian}) above can map onto a disordered single-particle tight-binding model on a Fock-space network with $\mathcal{N}$ sites. Setting the Fock states $\{|\mathbf{s}\rangle\}$ as the basis, we have
\begin{equation}\label{eq:hamiltonian_F}
        \frac{H_\mathrm{F}}{\hbar} =  \sum_{\mathbf{s}} \mathcal{E}_{\mathbf{s}}|\mathbf{s}\rangle\langle\mathbf{s}| + \sum_{\mathbf{s},{\mathbf{s}^{\prime}}} \mathcal{T}_{\mathbf{s}{\mathbf{s}^{\prime}}}|\mathbf{s}\rangle\langle\mathbf{s}^{\prime}|,
\end{equation}
where $\mathcal{E}_\mathbf{s}=\sum_{m,n}V_{mn}s^{mn}$ is the on-site energy for the basis $|\mathbf{s}\rangle$ and $\mathcal{T}_{\mathbf{s} \mathbf{s}^\prime}=\langle\mathbf{s}|H_{\mathrm R}|\mathbf{s}^{\prime}\rangle$ is the hopping strength between Fock states $|\mathbf{s}\rangle$ and $|\mathbf{s}^{'}\rangle$. As shown in Fig.~\ref{fig:1}{\bf b}, with the initial configuration $\mathbf{s}_0$ as the apex,  $\mathcal{N}$ Fock-space sites $\{\mathbf{s}\}$ can be sorted out as a 13-layer structure by the Hamming distance $d = 0, 2, \ldots$ 24 ($L$) wherein each layer includes $C^2(L/2,d/2)$ sites. Together with the connectivity provided by the hoppings $\{\mathcal{T}_{\mathbf{s}{\mathbf{s}^{\prime}}}\}$, the Fock-space Hamiltonian (\ref{eq:hamiltonian_F}) forms a quantum network with a double-cone structure. 

The experimental observation starts with a Fock state $|\mathbf{s}_0\rangle$ by exciting half of the qubits via $\pi$-pulses, which enables the probe of the largest photon-number-conserved sector with the Hilbert space dimension of $\mathcal{N}=2,704,156$. After suddenly opening interactions by tuning qubits and couplers, the system undergoes the out-of-equilibrium evolution governed by the engineered many-body Hamiltonian above (see Supplementary Section 3 for experimental sequences and calibrations). Finally, we extract the system dynamics with subsequent site-resolved simultaneous qubit readout after time $t$. The maximum evolution time $t_{\rm max}=1000~{\rm ns}$ is much longer than typical tunneling time $1/J_0\approx32~{\rm ns}$, enabling observation close to equilibration, while sufficiently short compared with qubit relaxation and dephasing times. Here, $J_0/2\pi\approx4.9~{\rm MHz}$, is the average of absolute values for all nearest-neighbor hoppings. For the detailed information about  error mitigation, decoherence effect, and finite-time effect, see Supplementary Sections 12, 13, and 14.

\section*{Wave packet dynamics in Fock space}
To describe quantum dynamics within the Fock-space network, a \emph{dynamical} radial probability distribution $\Pi(d,t)$ is introduced as
\begin{equation}\label{eq:distribution}
        \Pi(d, t) = \sum_{\mathbf{s}\in \{D(\mathbf{s}, \mathbf{s}_0)=d\}}|\langle\mathbf{s}|e^{-\frac{{i}H_{\mathrm{F}}t}{\hbar}}|{\mathbf{s}_0}\rangle|^2,
\end{equation}
which behaves as a wave packet and is interpreted as the probability that the many-body wave function appears at a Hamming distance $d$ away from the initial state $|\mathbf{s}_0\rangle$ at time $t$. In contrast with the experimentally inaccessible and static definition in terms of  eigenstates~\cite{Tomasi2021PRB}, $\Pi(d, t)$ quantitatively characterizes how an initial localized many-body wave function propagates in Fock space (Fig.~\ref{fig:1}{\bf c}), being experimentally feasible with growing system sizes. 

Examples of dynamics of $\Pi(d,t)$ for thermalization, localization and scarring are shown in Figs.~\ref{fig:2}{\bf a}, {\bf b} and {\bf c}, respectively. At $t=0$~ns, the initial state is a localized wave packet, thus $\Pi(0,0)= 1$. For thermalizing dynamics in the absence of disorder (Fig.~\ref{fig:2}{\bf a}), $|{\mathbf s}_0^T\rangle$ quickly and homogeneously spreads over the whole Fock space,  equilibrating near $L/2$ (Supplementary Section 10), where the density of states is maximum. In contrast, under the nonergodic dynamics ($V\approx 16 J_0$), the wave packet does not propagate to the large Hamming distance regime, staying around ($d<L/2$) its initial state, ultimately suggesting the localization in Fock space (Fig.~\ref{fig:2}{\bf b}). Interestingly, however, the scarred state $|{\mathbf s}_0^S\rangle$ shows a coherent periodic motion (Fig.~\ref{fig:2}{\bf c}) in Fock space. Due to the weak link between the scarred subspace and the huge thermal Hilbert space, the system exhibits slow thermalization~\cite{zhang2022ArXiv} (Supplementary Sections 5). The collective motion of twelve photons in real space can be reinterpreted as a Fock-space single-particle quantum walk~\cite{Yan2019Science, Braumuller2021NP}, where the wave packet behaves as a ballistic particle propagating along the Fock-space network, reflecting at the boundary at early times. However, in the long-time regime, it gradually escapes from this subspace,  diffusing in the entire Fock space, resulting in thermalization. For experimental results of few-body observables in real space, see Supplementary Sections 6, 7.

\begin{figure}
    \includegraphics[width=.95\columnwidth]{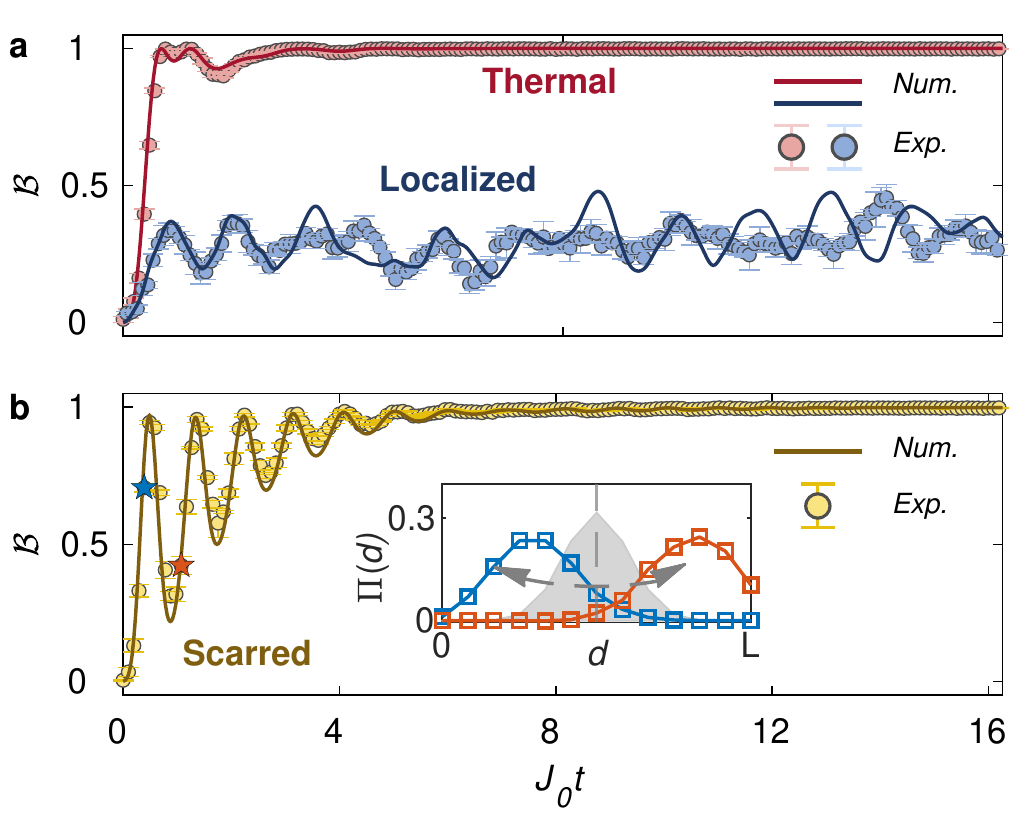}
    % \vspace{-0.6cm}
    \caption{
        {\bf Dynamics of Bhattacharyya distance ${\mathcal B}(t)$.} 
        {\bf a}, Dynamics of  ${\mathcal B}(t)$ for state $|{\mathbf s}_0^T\rangle$ at $V=0$ (red) and $V\approx{J_0}$ (blue). 
        {\bf b}, Dynamics of ${\mathcal B}(t)$ for scarred state $|{\mathbf s}_0^S\rangle$ at $V=0$, where the short-time oscillations and slow thermalization phenomena are observed due to the weak ergodicity breaking mechanism. Inset plots snapshots of experimental $\Pi(d)$ at $J_0t\approx0.4$ (blue) and $J_0t\approx1$ (red) for scarred dynamics, in contrast to the shaded background for the ergodic wave packet. Error bars stem from repetitions of measurements.
    }
    \label{fig:3}
\end{figure}

A generic many-particle system unaidedly thermalizes obeying the eigenstate thermalization hypothesis (ETH) \cite{Deutsch1991PRA,Rigol2008Nature}, which states that an ergodic system explores all allowed regions in phase space with the probability proportional to the space volume. Therefore, within the Fock-space picture, we can define an ergodic wave packet $\Pi^{\rm Erg.}(d)$ whose probability distribution is equal to the density of states (Supplementary Section 10). $\Pi^{\rm Erg.}(d)$ resembles a Gaussian distribution centered at $L/2$~(the gray region in the inset of Fig.~\ref{fig:3}{\bf{b}}). To describe the emergence of ergodicity or its breakdown, we introduce the Bhattacharyya distance
\begin{equation}
    {\mathcal B}(t)=\sum_{d}\sqrt{\Pi(d,t)\cdot\Pi^{\rm Erg.}(d)},
\end{equation}    
which quantifies the similarity between the dynamical wave packet $\Pi(d,t)$ and the ergodic one $\Pi^{\rm Erg.}(d)$. 

Figure~\ref{fig:3} shows the results of ${\mathcal B}(t)$. The prepared initial states are far from equilibrium and thus ${\mathcal B}(t=0)\approx0$. For $V=0$, the initial wave packet spreads over the entire space, indistinguishable from the ergodic ensemble $\Pi^{\rm Erg.}(d)$, thus $\mathcal{B}(t\to\infty)\to1$ for both thermal and scarred states. Albeit of final thermal fate, the weak ergodicity breaking mechanism leaves a slow-thermalization imprint on the scarred dynamics with periodic coherent oscillations (Fig.~\ref{fig:3}{\bf b}). In the deep localized phase ($V\approx 16J_0$), $\mathcal{B}(t)$ approaches a value far less than $1$, indicating that the wave packet restricts to a finite fraction of the Fock space, signifying the strong violation of ergodicity. A small $\mathcal{B}(t)$ means that only part of the Fock space actively contributes to the system dynamics, which can be instructive for developing a decimation scheme of the Hilbert space to efficiently simulate MBL systems with numerics~\cite{Aoki1980JPC,Pietracaprina2021AP}.

\begin{figure*} \includegraphics[width=1.8\columnwidth]{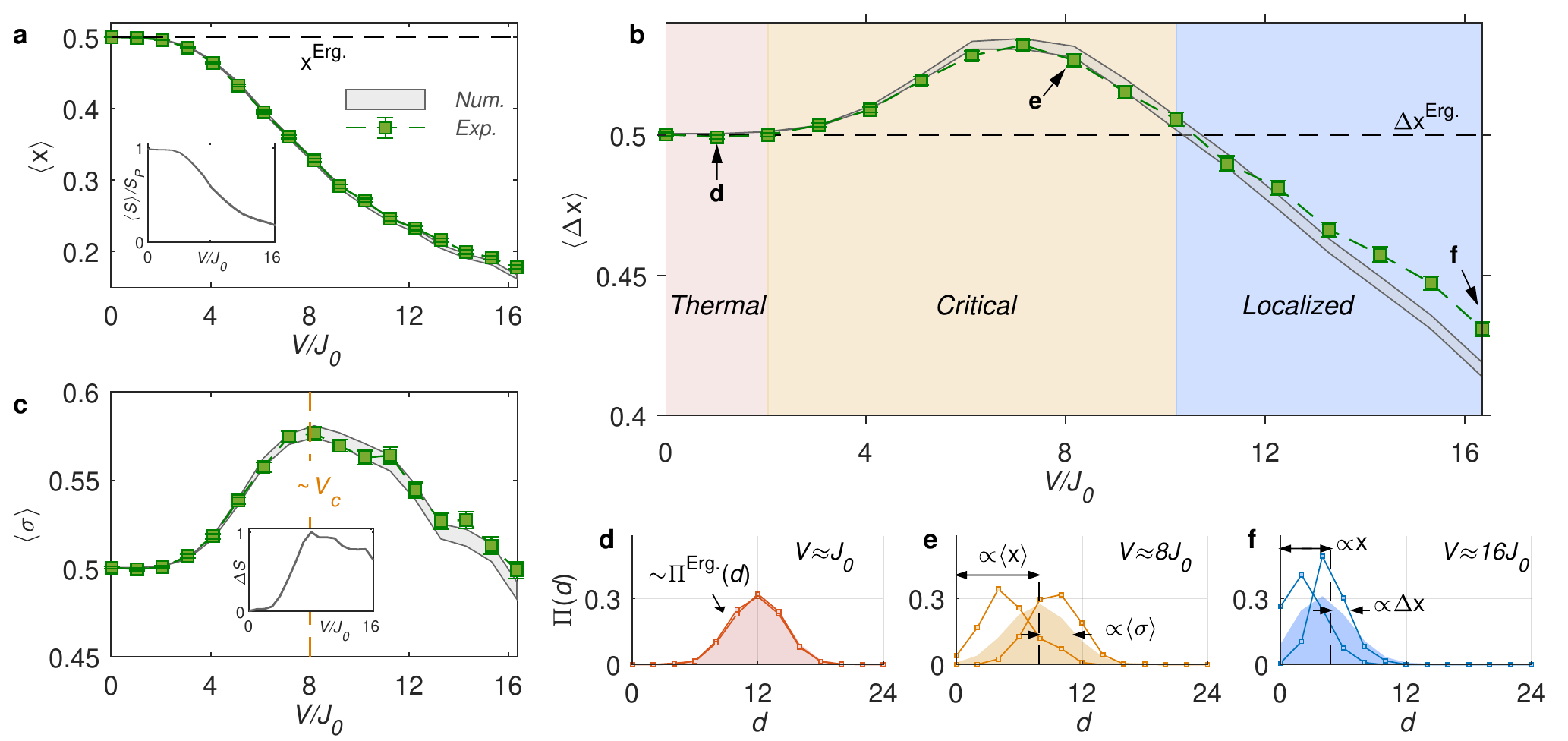}
    \caption{
    {\bf Signature of nonergodic transition in two dimensions.} 
    {\bf a}-{\bf c}, Disorder-averaged normalized displacement ${\rm \langle x\rangle}$, normalized width ${\rm \langle\Delta x\rangle}$, and system fluctuation ${\rm\langle\sigma\rangle}$ of wave packets at $t=1000$~ns~($J_0t\approx31$) as a function of disorder strength $V$, where the experimental data (green squares) are obtained by averaging over $400$ random disorder realizations in comparison with numerics (shades with boundary lines). Insets in {\bf a} and {\bf c} show simulation results for the disorder-averaged bipartite entanglement entropy $\langle S\rangle$ and its standard deviation $\Delta S$, exhibiting qualitatively similar behavior to ${\rm \langle x \rangle}$ and $\langle\sigma\rangle$, respectively. Error bars are the standard error of the statistical mean. {\bf d}-{\bf f}, Disorder-averaged wave packets (the shaded regions) for $V\approx J_0$ (red), $8 J_0$ (orange), and $16J_0$ (blue) over $400$ disorder realizations. Squares connected with lines in each panel are two representative realizations of $\Pi(d)$ at the corresponding disorder strength.  
    }
    \label{fig:4}
\end{figure*}

\section*{signature of two-dimensional nonergodic transition}
In addition to characterizing many-body dynamics over different quantum phases, more importantly, our Fock-space approach provides a \emph{scalable} way being capable of exploring critical phenomena. Quantum critical behavior near the MBL transition currently lacks consensus \cite{Khemani2017PRX,Suntajs2020PRE} and its locating method typically relies on \emph{global} diagnostics (e.g., bipartite EE and IPR)~\cite{Jonas2014PRL, Luitz2015PRB, Evers2008RMP, Luitz2014PRB,Nicolas2019PRL,Suntajs2020PRB,Khemani2017PRX}, which are experimentally and numerically challenging for large systems (Supplementary Section 8, 11). By analyzing properties of the wave packet in detail, we find three regimes for the MBL \emph{crossover} in our finite-sized 2D system. 

We utilize $\Pi(d)$ in the long-time limit ($t=1000$~ns; $J_0t\approx31$) to characterize the transition from the thermal to the MBL-like phase (whose approach to the thermodynamic limit in the latter requires confirmation). We introduce the normalized displacement ${\rm x}=\frac{1}{L}\sum_{d}{d}\Pi(d)$ and the normalized width ${\rm \Delta x}= \frac{\sqrt{L-1}}{L}\sqrt{\sum_{d}{d^2}\Pi(d) - [\sum_{d}{d}\Pi(d)]^2}$ to quantify the properties of the wave packet. For an ergodic wave packet, ${\rm x}^{\rm Erg.}=0.5$ and ${\rm \Delta x}^{\rm Erg.}=0.5$ (Supplementary Section 10).

Figure~\ref{fig:4}{\bf a}  displays the disorder-averaged displacement $\langle{\rm x}\rangle$ as a function of disorder strength $V$, where $\langle\cdots\rangle$ denotes the average over $k=400$ combinations of random disorder realizations (Supplementary Section 9). Notably, ${\rm x}$ is closely related to the local real-space autocorrelation function, implying a preservation of a local information for ${\rm x}<0.5$~\cite{Hauke2015PRB,Smith2016np, Guo2021PRL}. Deep in the thermal phase with $V\to0$, $\langle{\rm x}\rangle\to0.5$, in quantitative agreement with ${\rm x}^{\rm Erg.}$. As disorder strength grows, $\langle{\rm x}\rangle$ decreases monotonically, meaning that the propagation of the wave packet in Fock space is hindered by disorder, and the initial local information is preserved, suggesting strong evidence of nonergodicity. Its behaviors are compatible with the numerical results of bipartite entanglement entropy $\langle S\rangle/S_P$~(the inset of Fig.~\ref{fig:4}{\bf a}). Here, $S_P=0.5[{\rm ln}(2)L-1]$ is the Page value for random states in Fock space~\cite{Page1993}.  

As shown in Fig.~\ref{fig:4}{\bf b}, the second-order quantity $\langle{\rm \Delta x}\rangle$, without real-space correspondence, exhibits much richer features and allowing the identification of three typical regimes for the disorder-induced transition from the thermal phase to the MBL-like phase. In the weak disorder regime $V \lesssim 2J_0$, we find a plateau of $\langle{\rm \Delta x}\rangle = {\rm \Delta x}^{\rm Erg.}$, indicating a thermal phase, where wave packets almost coincide perfectly with the ideal thermal one $\Pi^{\rm Erg.}(d)$ (Fig.~\ref{fig:4}{\bf d}). A critical regime of $2J_0 \lesssim V\lesssim 10J_0$ is identified by an anomalous increase of wave packet width with $\langle{\rm \Delta x}\rangle \gtrsim {\rm \Delta x}^{\rm Erg.}$, which shows a good agreement with the anomalous slow-relaxation regime reported in Ref.~\cite{BordiaPRX2017}. Take $V\approx 8J_0$ for example~(Fig.~\ref{fig:4}{\bf e}): the wave packet in this regime exhibits a broader distribution with an enhanced sample-to-sample fluctuation for the displacement ${\rm x}$. At strong disorder regime $V\gtrsim10J_0$, the width plummets below ${\rm \Delta x}^{\rm Erg.}$, implying localization in Fock space. Despite the narrower wave packets with the smaller displacement, the displacement fluctuations from sample to sample are still larger than that in the thermal phase (Fig.~\ref{fig:4}{\bf f}). 

To locate the critical disorder, we use the fluctuation $\sigma= \frac{\sqrt{L-1}}{L}\sqrt{\sum_{d}{d^2}\Pi(d) - \langle{\rm x}\rangle^2}$, the width of disorder-averaged wave packet~(see Fig.~\ref{fig:4}{\bf e}), as the diagnostic. It comes from sample-to-sample randomness, including both displacement fluctuations and width variations. Enhanced fluctuations are observed as shown in Fig.~{\ref{fig:4}}{\bf c}. A putative critical disorder $V_c\approx 8 J_0$ is estimated by the peak position of $\langle\sigma\rangle$. Remarkably, the numerics for $\Delta S$ (the standard deviation of bipartite EE) predicts the same value (the inset of Fig.~\ref{fig:4}{\bf c}), validating the $V_c$ estimated within our approach.

The observed physical picture above can be explained by a dilute thermal bubbles model~\cite{Tomasi2021PRB}, which  establishes the connection between thermal avalanche theory~\cite{Roeck2017PRB, Potirniche2019PRB} and the renormalization group approach~\cite{Goremykina2019PRL,Dumitrescu2019PRB} for understanding of MBL transition. Currently, the existence of MBL in a 2D system is still widely debated~\cite{Foo2022Arxiv}. 
Despite numerical~\cite{Wahl2018NP,Kshetrimayum2020,eveniaut2020PRR,Chertkov2021PRL}  and experimental~\cite{Choi2016science, BordiaPRX2017} evidences, thermal avalanche theory argues on the instability of a 2D MBL phase due to thermal bubbles occurring in rare regions of locally weak disorders. A final conclusion requires proper scaling and the verification of the delocalization mechanism in the presence of such bubbles, which is beyond the scope of this work. 

\section*{conclusion and outlook}
We experimentally demonstrate a novel protocol of exploring many-body physics from a fresh perspective -- Fock space. By mapping the out-of-equilibrium many-body dynamics onto a wave packet propagating on a Fock-space network, we provide a clear view of the thermalization and its breakdown in Fock space, described by the dynamical trajectory of the wave packet and the statistical emergence of an ergodic ensemble. In Fock space, key features of MBL, inhibition of wave packet propagation, are observed dynamically, and many-body scarring is identified by short-time oscillations. Besides that, our protocol also allows us to experimentally capture the elusive quantum critical behaviors across the finite-size MBL transition in two dimensions. A three-regime picture of such transition and a critical disorder $V_c$ are quantitatively identified experimentally, which is challenging for traditional real-space observations.

Methodologically, our protocol provides a simple yet effective way to quantitatively reveal the nature of MBL transition, even for critical behaviors of higher-dimensional systems. It is worth stressing that unlike experimentally and numerically prohibitive bipartite EE and IPR in large systems, our protocol is scalable, platform-independent, and robust to readout errors~(Supplementary Section 12), making it a universal experimental playground to solve contentious questions in future larger devices. 

Although our experiments have already provided significant insights and implications for understanding MBL in Fock space, conclusive arguments on open questions involved in this work, such as the stability of MBL in higher dimensions~\cite{Foo2022Arxiv} and the role of rare regions near the critical point~\cite{Roeck2017PRB, Luitz2016PRB} need further investigations. Going forward, it will be fascinating to develop a finite-size scaling method based on our protocol, the application of which to larger quantum devices, surpassing sizes amenable to classical computations, has a larger chance to settle the ongoing conundrum.  

\section*{Author contributions}
Q.G. and L.Y. proposed the idea.,
Y.Y., L.X. and Z.B. conducted the experiment and analyzed the data  under the supervision of Q.G. and H.W., 
Z.B., Z.G., and Y.-F.Y. performed the numerical simulation under the supervision of Q.G., C.C. and L.Y., 
H.L. and J.C. fabricated the device under the supervision of H.W.,
Q.G., L.Y., R.M. and H.W. co-wrote the manuscript,
and S.-Y.Z. supervised the whole project.
All the authors contributed to the experimental setup, discussions of the results and development of the manuscript.

\section*{acknowledgments}
%R.M.~acknowledges support from the NSFC Grants No.~U1930402, 12111530010, 12222401, and 11974039.
The device was fabricated at the Micro-Nano Fabrication Center of Zhejiang University. We acknowledge support from the National Natural Science Foundation of China (Grant Nos. 92065204, U20A2076, 12274368, U1930402, 12111530010, 12222401, 11974039,  12174167 and  12047501), and the Zhejiang Province Key Research and Development Program (grant no. 2020C01019). L.Y. is also supported by National Key Research and Development Program of China (Grant No. 2022YFA1404203) and the Fundamental Research Funds for the Central Universities.

\bibliographystyle{naturemag}
%\bibliography{mainRef.bib}

\clearpage
%\begin{document}

\beginsupplement
\begin{center}
{\bf {\Large Supplementary Information}}
\end{center}

\section{Experimental setup}
\begin{figure*}
  \includegraphics[width=1.8\columnwidth]{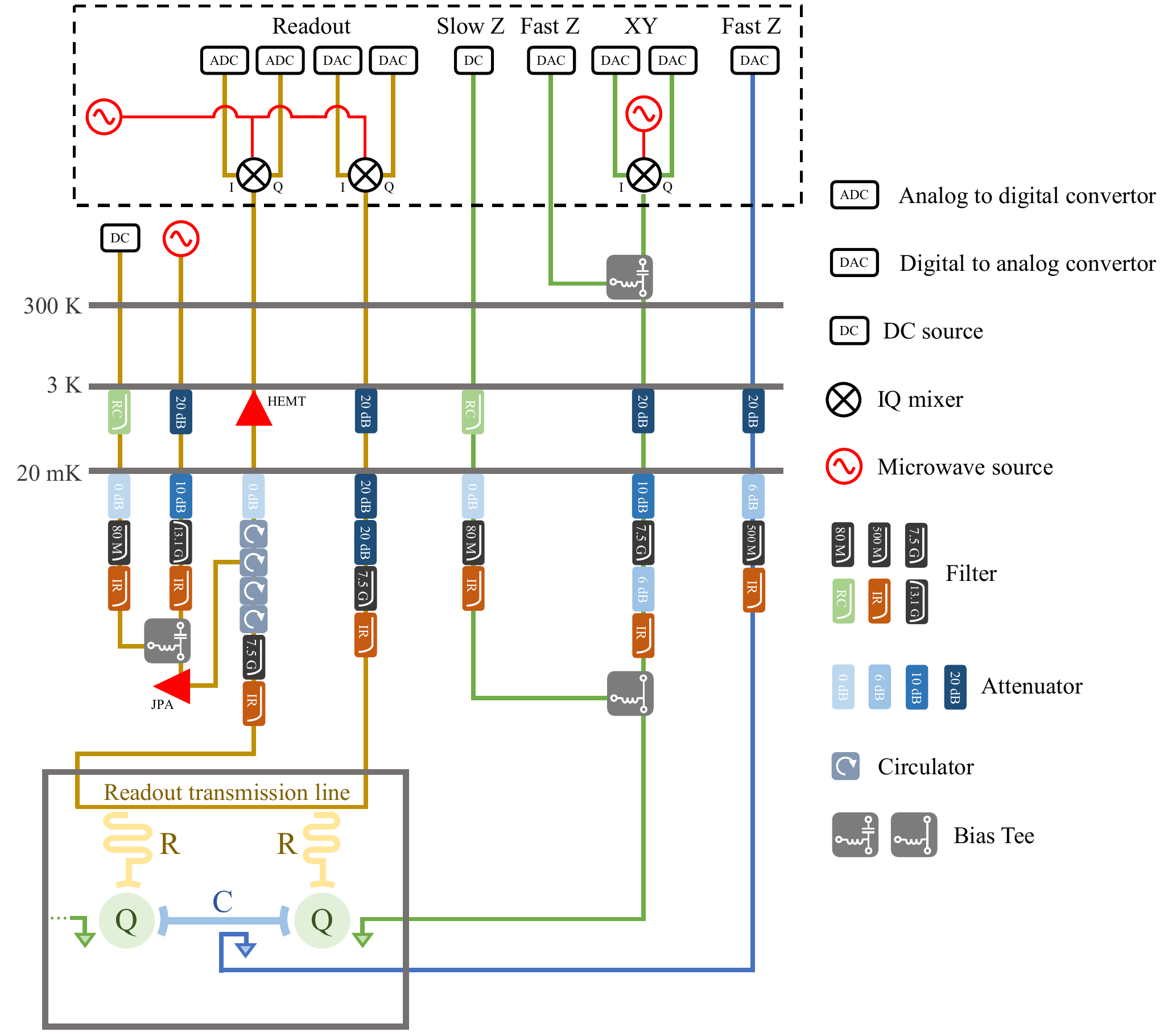}
   \caption{
   {\bf Experimental setup.}
   The cartoon diagram schematically shows the connection of room-temperature electronics (DACs, ADCs, and IQ mixers) and cryogenic wirings (cables, attenuators, filters, circulators, and amplifiers) for controlling and measuring our device. The layout in the left-lower gray box represents the device, where Q, C, and R indicate the qubit, coupler, and readout resonator, respectively. Different stages of DR, including mixing chamber~(20 mK), $3$~K plate and $300$~K plate are indicated by horizontal thick gray lines, where attenuators, filters, circulators and amplifiers are mounted. The top dashed rectangular box illustrates the room-temperature electronics for generating control signals and demodulating readout signals.
   }
   \label{Figure:ExpSetup}
\end{figure*}

As shown in Fig.~1{\bf a} of the main text, our experiment is carried out on a two-dimensional (2D) flip-chip superconducting quantum processor, which integrates $36$ transmon qubits and $60$ tunable couplers~\cite{Yan2018PRAp}. The former form a $6\times6$ square lattice, and the latter provide the nearest-neighbor connectivity. A diagrammatic sketch of on-chip control circuits for two qubits and a coupler is shown in the gray box of Fig.~\ref{Figure:ExpSetup}. Each qubit owns an individual control line, which enables the microwave (XY) control for the single-qubit rotation and the flux (Z) control for modulating its frequency. Meanwhile, each qubit is capacitively coupled to a readout resonator for its state discrimination. Similarly, each coupler owns an individual flux control line for tuning frequency. In this way, the coupling strength between the two connected qubits by the coupler can be effectively tuned as needed.

The experimental setup for the control and the measurement of this multi-qubit device is shown in Fig.~\ref{Figure:ExpSetup}. The device is loaded inside a dilution refrigerator~(DR) with a base temperature of  $\sim20$~mK. The control pulses, such as XY pulses, fast (slow) Z pulses, and microwave readout pulses are generated by room-temperature electronics (dashed box in Fig.~\ref{Figure:ExpSetup}), which are transmitted to the device after a series of attenuating, filtering at different cold stages in the DR. The digital to analog converter (DAC) channels are used individually for the fast Z controls of qubits and couplers, and in pairs for generating XY control pulses or readout pulses via frequency mixing with local oscillator signals. In addition, there is a low-noise slow Z (DC) control channel for biasing qubit to its idle frequency. For the frequency-multiplexed qubit readout, every nine qubits share a common readout transmission line, and the output readout signals from the device are amplified sequentially by a Josephson parametric amplifier (JPA) at the mixing chamber and a high electron mobility transistor (HEMT) amplifier at the 3K stage. At room temperature, readout signals are further amplified by room-temperature amplifiers and mixed down via the IQ mixer module. Finally, qubit state information is extracted with a demodulation process on analog-to-digital converters (ADCs).

\section{Device performance}
   The device is fabricated using the flip-chip recipe (see Ref.~\cite{zhang2022Nature} for more fabrication details). In current experiment, we utilize up to $24$ ($4\times6$) qubits and $38$ couplers, but we collect the performance for all 36 qubits in Table~\ref{Tab:1}. It is outstanding that the average energy relaxation time $T_{1}$ is more than 120 $\mu$s in our device. In addition, the dephasing time measured with spin echo (Hahn echo) experiment is $\sim$23 $\mu$s. Both of them are much longer than the maximum evolution time ($t_{\rm max}=$1000~ns) in our experiment, and thus make it possible to observe the coherent quantum many-body dynamics before decoherence effects are substantial. Average single-qubit gate fidelities measured by simultaneous randomized benchmarking are $\sim0.9968$ for 36 qubits (see Table~\ref{Tab:1}), and $\sim0.9978$ for the 24 qubits used in our experiment.   
  
\begin{table*}[t] 
    \renewcommand\tabcolsep{12.0pt}
    \begin{tabular}{ccccccccc}
      \hline
      \hline
                    Qubit & $\omega_{(m,n)}/2\pi$   & $T_{1, {(m,n)}}^{(\mathrm{idle})}$ & $T_{1,(m,n)}^\mathrm{(inte.)}$ & $T^{\ast}_{2, (m,n)}$  & $T^{SE}_{2, (m,n)}$   & $F_{0, (m,n)}$   & $F_{1, (m,n)}$   & $e_\mathrm{sq}$ \\
                     Index & (GHz)   & ($\mu$s)& ($\mu$s)&  ($\mu$s) &  ($\mu$s)  &    &   &  ($\%$)\\
      \hline
      $Q_{(1,1)}$      & 4.630     &$\sim$169  &$\sim$147 &$\sim$3.6    & $\sim$10.7    & 0.973     & 0.950     & 0.20 \\
      $Q_{(1,2)}$      & 4.780     &$\sim$117  &$\sim$114 &$\sim$8.8    & $\sim$27.3    & 0.988     & 0.962     & 0.40 \\
      $Q_{(1,3)}$      & 4.881     &$\sim$130  &$\sim$119 &$\sim$7.0    & $\sim$19.8    & 0.979     & 0.946     & 0.53 \\
      $Q_{(1,4)}$      & 4.704     &$\sim$132  &$\sim$132 &$\sim$3.9    & $\sim$13.3    & 0.976     & 0.946     & 0.58 \\
      $Q_{(1,5)}$      & 4.597     &$\sim$96   &$\sim$100 &$\sim$4.7    & $\sim$21.4    & 0.974     & 0.893     & 0.17 \\
      $Q_{(1,6)}$      & 4.509     &$\sim$150  &$\sim$114 &$\sim$5.2    & $\sim$20.0    & 0.970     & 0.915     & 0.14 \\
      $Q_{(2,1)}$      & 4.368     &$\sim$160  &$\sim$117 &$\sim$5.0    & $\sim$21.6    & 0.967     & 0.923     & 0.27 \\
      $Q_{(2,2)}$      & 4.826     &$\sim$163  &$\sim$90 &$\sim$2.8    & $\sim$9.45    & 0.975     & 0.928     & 0.22 \\
      $Q_{(2,3)}$      & 4.752     &$\sim$132  &$\sim$87 &$\sim$6.2   & $\sim$21.7    & 0.986     & 0.952     & 0.13 \\
      $Q_{(2,4)}$     & 4.679     &$\sim$132  &$\sim$118 &$\sim$3.8    & $\sim$13.8    & 0.979     & 0.899     & 0.33 \\
      $Q_{(2,5)}$     & 4.643     &$\sim$138  &$\sim$128 &$\sim$3.0   & $\sim$13.4    & 0.976     & 0.926     & 0.22 \\
      $Q_{(2,6)}$     & 4.471     &$\sim$108  &$\sim$139 &$\sim$10.6     & $\sim$36.9    & 0.972     & 0.912     & 0.20 \\ 
      $Q_{(3,1)}$     & 4.542     &$\sim$119  &$\sim$150 &$\sim$7.3    & $\sim$30.9    & 0.978     & 0.888     & 0.15 \\
      $Q_{(3,2)}$     & 4.791     &$\sim$142  &$\sim$120 &$\sim$6.0    & $\sim$20.7    & 0.953     & 0.947     & 0.33 \\
      $Q_{(3,3)}$     & 4.615     &$\sim$110  &$\sim$136 &$\sim$3.6   & $\sim$10.4    & 0.966     & 0.898     & 0.28 \\
      $Q_{(3,4)}$     & 4.722     &$\sim$133  &$\sim$157 &$\sim$8.7   & $\sim$28.7    & 0.989     & 0.970     & 0.35 \\
      $Q_{(3,5)}$     & 4.666     &$\sim$139  &$\sim$160 &$\sim$4.0    & $\sim$14.1    & 0.982     & 0.954     & 0.59 \\
      $Q_{(3,6)}$     & 4.492     &$\sim$106  &$\sim$130 &$\sim$9.0    & $\sim$33.7    & 0.969     & 0.863     & 0.45 \\
      $Q_{(4,1)}$     & 4.387     &$\sim$141  &$\sim$114 &$\sim$7.5    & $\sim$20.7    & 0.974     & 0.936     & 0.24 \\
      $Q_{(4,2)}$     & 4.432     &$\sim$155  &$\sim$152 &$\sim$6.6    & $\sim$19.4    & 0.969     & 0.949     & 0.38 \\
      $Q_{(4,3)}$     & 4.522     &$\sim$152  &$\sim$132 &$\sim$5.2    & $\sim$15.0    & 0.981     & 0.942     & 0.28 \\
      $Q_{(4,4)}$     & 4.761     &$\sim$157  &$\sim$127 &$\sim$7.3    & $\sim$24.7    & 0.987     & 0.942     & 0.13 \\
      $Q_{(4,5)}$     & 4.803     &$\sim$154  &$\sim$112 &$\sim$6.3    & $\sim$23.5    & 0.990     & 0.948     & 0.12 \\
      $Q_{(4,6)}$     & 4.560     &$\sim$121  &$\sim$118 &$\sim$5.3    & $\sim$21.1    & 0.988     & 0.952     & 0.16 \\ 
      $Q_{(5,1)}$     & 4.813     &$\sim$114  &$\sim$133 &$\sim$8.8    & $\sim$17.5    & 0.917     & 0.873     & 0.40 \\ 
      $Q_{(5,2)}$     & 4.652     &$\sim$127  &$\sim$120 &$\sim$7.6    & $\sim$23.5    & 0.957     & 0.872     & 0.42 \\ 
      $Q_{(5,3)}$     & 4.686     &$\sim$89   &$\sim$120 &$\sim$7.4    & $\sim$31.9    & 0.942     & 0.862     & 0.29 \\ 
      $Q_{(5,4)}$     & 4.715     &$\sim$103  &$\sim$89 &$\sim$14.1    & $\sim$58.0    & 0.953     & 0.916     & 0.73 \\ 
      $Q_{(5,5)}$     & 4.604     &$\sim$104  &$\sim$99 &$\sim$5.8     & $\sim$18.1    & 0.974     & 0.886     & 0.36 \\ 
      $Q_{(5,6)}$     & 4.751     &$\sim$131  &$\sim$102 &$\sim$10.5     & $\sim$42.1    & 0.972     & 0.947     & 0.28 \\ 
      $Q_{(6,1)}$     & 4.764     &$\sim$151  &$\sim$150 &$\sim$6.4    & $\sim$19.8    & 0.988     & 0.944     & 0.73 \\ 
      $Q_{(6,2)}$     & 4.403     &$\sim$118  &$\sim$102 &$\sim$3.1    & $\sim$9.7    & 0.921     & 0.886     & 0.23 \\ 
      $Q_{(6,3)}$     & 4.462     &$\sim$115  &$\sim$155 &$\sim$4.9    & $\sim$15.5    & 0.960     & 0.900     & 0.26 \\ 
      $Q_{(6,4)}$     & 4.838     &$\sim$127  &$\sim$129 &$\sim$8.3    & $\sim$35.1    & 0.973     & 0.924     & 0.17 \\ 
      $Q_{(6,5)}$     & 4.806     &$\sim$116  &$\sim$129 &$\sim$9.4    & $\sim$30.5    & 0.969     & 0.944     & 0.62 \\ 
      $Q_{(6,6)}$     & 4.858     &$\sim$121  &$\sim$101 &$\sim$7.6    & $\sim$37.2    & 0.963     & 0.938     & 0.22 \\ 
      \hline
      mean      & 4.652    & $\sim$130  &$\sim$123 &$\sim$6.5 & $\sim$23.1   & 0.970     & 0.925     & 0.32 \\
%       mean      & 4.631    & $\sim$135  &$\sim$123 &$\sim$6.5 & $\sim$20.5   & 0.977     & 0.931     & 0.22 \\
      \hline
      \hline
    \end{tabular}
    \caption{{\bf Device performance.} 
    \label{Tab:1}$\omega_{(m,n)}$ is the idle frequency of $Q_{(m,n)}$ where single-qubit XY pulses are applied for preparing initial Fock states. $T_{1,(m,n)}^{(\mathrm{idle})}$  and $T_{1,(m,n)}^\mathrm{(inte.)}$  are the energy relaxation times of $Q_{(m,n)}$ measured at its idle frequency $\omega_{(m,n)}$ and the interaction frequency ($\omega_I/2\pi\approx$4.57~GHz), respectively. $T^{\ast}_{2, {(m,n)}}$ and $T^{SE}_{2, {(m,n)}}$ are the ramsey and spin echo dephasing times of  $Q_{(m,n)}$ measured at $\omega_{(m,n)}$. The readout fidelities are characterized by the measured probability when $Q_{(m,n)}$ is prepared in state $|0\rangle$ ($|1\rangle$), labeled by $F_{0,{(m,n)}}$ ($F_{1,{(m,n)}}$). Single-qubit gate errors $e_\mathrm{sq}$ are measured with randomized benchmarking (RB) on 36 qubits simultaneously. We note that such average $e_\mathrm{sq}$ is about 0.22\% for the simultaneous RB on the 24 actively used qubits in the experiment. 
    } 
    \label{T1}
\end{table*}

\section{Experimental calibrations}
\begin{figure}
  \includegraphics[width=0.9\columnwidth]{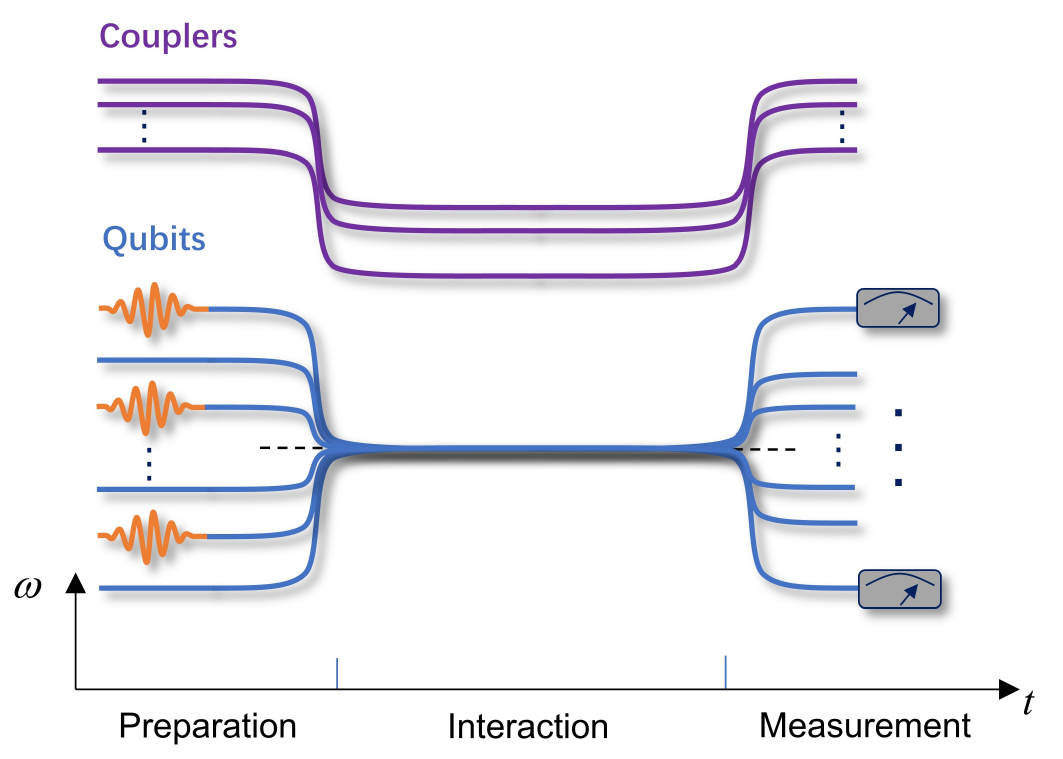}
   \caption{
   {\bf Pulse sequence.}
   There are three steps (preparation, interaction and measurement) for the experimental pulse sequence. $\omega$ axis indicates the frequency domain of qubits and couplers. The $\pi$ pulses for exciting qubits are denoted by orange Gaussian-shape envelopes. The black dashed line represents the interaction frequency $\omega_{I}/2\pi=4.57$~GHz where qubits interact with each other when disorders are absent. 
   }
   \label{Figure:sequence}
\end{figure}

We use quench protocol to investigate many-body dynamics. As shown in Fig.~\ref{Figure:sequence}, the experimental sequence includes three steps: state preparation, interaction and measurement. For the initial state preparation, all qubits and couplers stay at their idle points, where half of qubits are excited to $|1\rangle$ states by $\pi$ pulses (orange Gaussian-shape waveforms). Then, all qubits are suddenly tuned to their interaction frequencies via fast Z pulses. At the same time, all couplers are biased for realizing target coupling strengths in the same way. After an evolution time $t$, all couplers are tuned back to idle points and all qubits are tuned to their respective readout frequencies for state measurements.

\begin{figure}[b]
  \includegraphics[width=0.9\columnwidth]{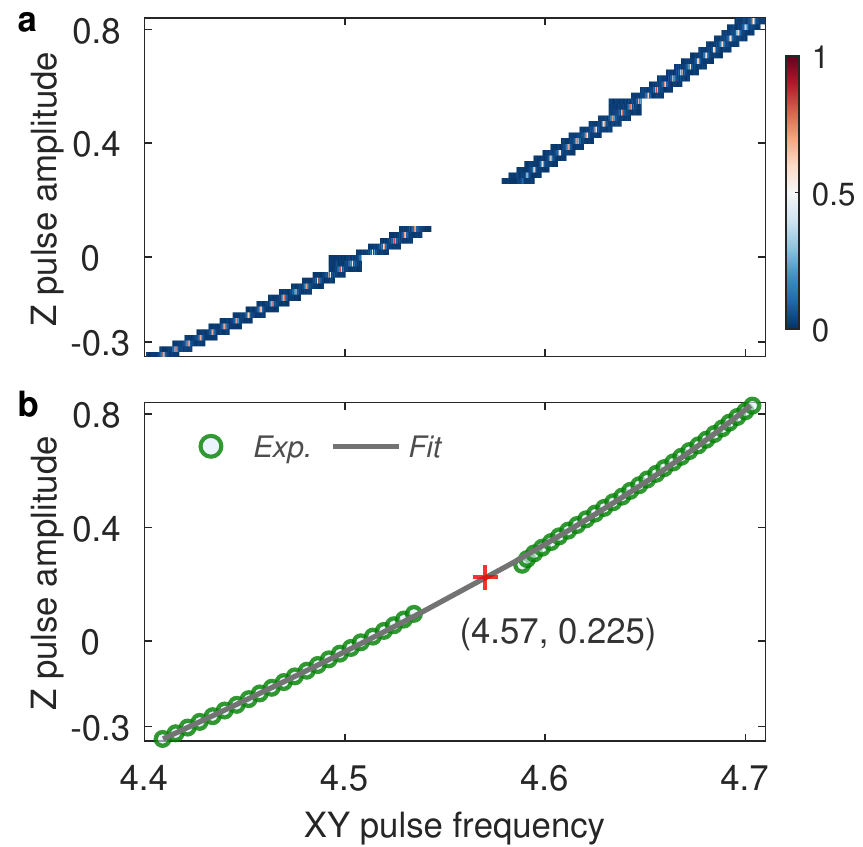}
   \caption{ {\bf Frequency calibrations: Z pulse amplitude versus qubit frequency.}
   {\bf a}, The raw data of the spectroscopy measurements for qubit $Q_{(4,6)}$ when all other qubits are tuned to $\omega_I$ and all the couplers are tuned to the target coupling configuration for many-body interactions. {\bf b}, The polynomial fitting (gray line) for the experimental spectroscopy data (green circles) extracted from {\bf a}.
  }
   \label{Figure:specAuto}
\end{figure}

In our experiment, fast Z pulses play a substantial role in tuning frequencies (disorder strength), which makes their calibrations more important. Here, we take qubit $Q_{(4,6)}$ as an example to show its calibrations. To minimize the crosstalk effect on final multi-qubit experiments, we keep all other qubits stay near the interaction frequency $\omega_I$ and all couplers stay on target coupling strength when we use spectroscopy experiment to map the relationship between qubit frequency and Z pulse amplitude. The measured spectroscopy data is shown in Fig.~\ref{Figure:specAuto}{\bf a}. There are two branches above and below $\omega_I$ due to the avoided crossing. With a polynomial fitting of these two branches (see Fig.~\ref{Figure:specAuto}{\bf b}), we could extract a function of Z pulse amplitude versus qubit frequency ranging from 4.4 GHz to 4.7 GHz. 

 %%%%%%%%%%%%%%%%%%%%%%%%%%%%%%%%%%%%%%%%%%%%%%%%%%%%%%%%%%%%%%%%%%%%%%%%
\section{Effective Hamiltonian}
Equation~(1) in the main text is the effective Hamiltonian of our model and its tunable interactions  between nearest-neighbor (NN) qubit pairs is enabled by the adjacent couplers. Here, we show the derivation of effective Hamiltonian from its original configurations including couplers. At first, the full Hamiltonian ($\hbar=1$) in a general configuration is written as
\begin{equation}
  \begin{aligned}
    H_{0}=& \sum_{m} \omega_{m} \sigma_{m}^{+}\sigma_{m}^{-}+\sum_{j} \omega_{j} \tau_{j}^{+}\tau_{j}^{-} \\
    &+\sum_{m,j} g_{mj}(\sigma_{m}^{+} \tau_{j}^{-}+\sigma_{m}^{-} \tau_{j}^{+}) \\
    &+\sum_{m,n}g_{mn}(\sigma_{m}^{+} \sigma_{n}^{-}+\sigma_{m}^{-} \sigma_{n}^{+}),
    \end{aligned}
\end{equation}
where $g_{mn}$ ($g_{mj}$) is the coupling strength between qubit $Q_m$ and qubit $Q_n$ (qubit $Q_m$ and coupler $C_j$), $\sigma_m^{\pm}$ and $\tau_j^{\pm}$ are the raising (lowering) operators for qubit $Q_m$ and coupler $C_j$, respectively. The long-range couplings are very small, thus, we only consider the qubit-qubit couplings for the qubit pairs separated with the distance $R_{mn} = |\mathbf{r}_m-\mathbf{r}_n| \leq \sqrt{2}a_0$, where $a_0$ is the lattice constant of 2D qubit lattice and $\mathbf{r}_{m}$, $\mathbf{r}_n$ are the positions of  $Q_m$ and $Q_n$, respectively. 

In the regime of $g_{mj}$ ($g_{nj}$) $>g_{mn}>0$, $\Delta_{m(n)}=\omega_{m(n)}-\omega_j<0$, and $g_{m(n)j}\ll|\Delta_{m(n)}|$, such a Hamiltonian can be applied with the Schrieffer-Wolff transformation~\cite{Bravyi_2011}
\begin{equation}
U= \sum_{R_{mn}=a_0}  e^{\frac{g_{mj}}{\Delta_m}\left(\sigma_m^+\tau_j^--\sigma_m^-\tau_j^+\right)+\frac{g_{nj}}{\Delta_n}\left(\sigma_n^+\tau_j^- -\sigma_n^-\tau_j^+\right) }
\end{equation}
to the first order of $g_{m(n)j}/\Delta_{m(n)}$. Then, we have the transformed Hamiltonian as
\begin{equation}
  \begin{aligned}
    {H} = \sum_m  \omega_{m}^{\prime} \sigma_{m}^{+}\sigma_{m}^{-}+ \sum_{m,n}J_{mn}(\sigma_{m}^{+} \sigma_{n}^{-}+\mathrm{h.c.}),
   \end{aligned}
\end{equation}
with 
\begin{equation}
    \begin{aligned}
      \omega_{m}^{\prime} = &  \omega_{m}+\frac{g_{mj}^2}{\Delta_{m} } , \\
        J_{mn}   = &  \frac{g_{mj}g_{nj}}{\Delta} + g_{mn}  , \\
        \frac{2}{\Delta} = & \frac{1}{\Delta_{mj}}+\frac{1}{\Delta_{nj}} .
    \end{aligned}
\end{equation}
In the rotating frame of $\omega_I$, we can get the effective Hamiltonian in real space as
\begin{equation}
  \label{eq:H_F}
  \begin{aligned}
    {H}_{\rm R} = \sum_m  V_{m} \sigma_{m}^{+}\sigma_{m}^{-}+ \sum_{m,n}J_{mn}(\sigma_{m}^{+} \sigma_{n}^{-}+\mathrm{h.c.}),
   \end{aligned}
\end{equation}
where $V_m=\omega_m^{\prime}-\omega_I$. Equation (1) of the main text is just the 2D form of Equation~(\ref{eq:H_F}) and $J_{mn}$ includes all the NN couplings ($\chi_m$, $\gamma_n$) and cross couplings ($g_x$). 

\begin{figure}[ht]
  \includegraphics[width=0.8\columnwidth]{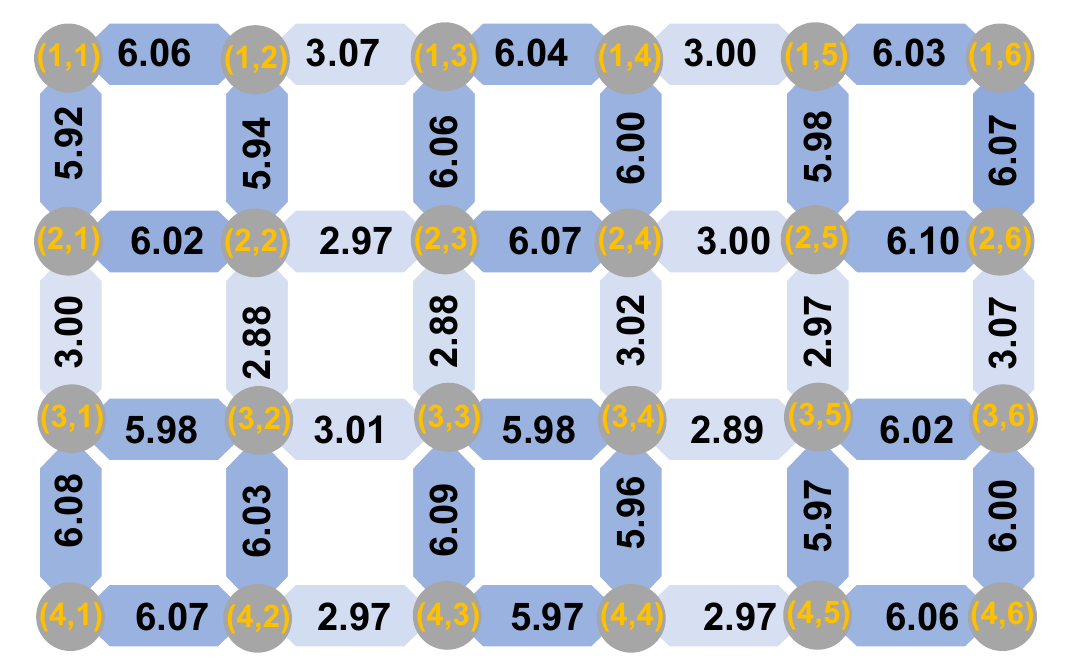}
   \caption{
   {\bf Experimentally measured nearest-neighbor couplings}.
   Gray circles represent qubits, and the chamfered rectangles connecting them denote the nearest-neighbor couplings with values for $-J_{mn}/2\pi$ (MHz) listed.
   }
   \label{Figure:NN_couplings}
\end{figure}

\begin{figure}[b]
  \includegraphics[width=0.9\columnwidth]{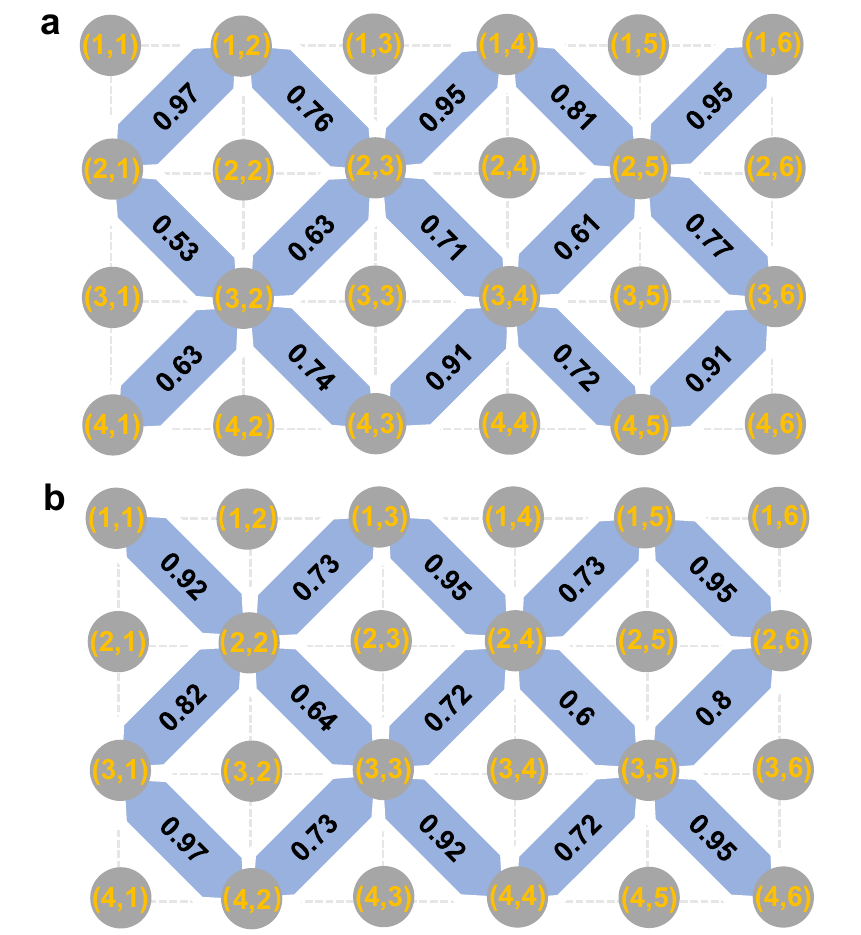}
   \caption{
   {\bf Experimentally measured  cross couplings.} 
   Gray circles represent qubits, and the chamfered rectangles connecting them denote the measured cross couplings with their values listed (MHz).
   }
   \label{Figure:cross_couplings}
\end{figure}

In order to realize a 2D SSH model with many-body scarring, we set $J_o/2\pi=2J_e/2\pi=-6~$MHz and $g_{\mathrm{x}}/2\pi=0.9$~MHz in our model, which are used for the numerical simulations. Effective coupling strengths $\{J_{mn}\}$ measured by two-qubit photon swapping experiments at $\omega_I/2\pi=4.57$~GHz are illustrated in Fig.~\ref{Figure:NN_couplings} for NN couplings and Fig.~\ref{Figure:cross_couplings} for cross couplings. We choose $V_{m}$ from a uniform random distribution $[-V, V]$ to mimic a fully disordered system. $V_{m}$ can be tuned individually by adjusting $Q_m$'s frequency without noticeably altering  $\{J_{m n}\}$.

%%%%%%%%%%%%%%%%%%%%%%%%%%%%%%%%%%%%%%%%%%%%%%%%%%%%%%
\section{Mechanism of 2D many-body scarring }
Inspired by the scarring phenomenon in the 1D XY model~\cite{zhang2022ArXiv}, here, we consider a 2D SSH lattice constructed by a set of tetramers. The intra- and inter-tetramer couplings are denoted by $J_\mathrm{o}$ and $J_\mathrm{e}$, respectively. 
To understand the scarring phenomenon, we begin to focus on a single tetramer with a square geometry that includes four qubits. In the half-filling condition, a special kind of entangled eigenstates is given by
\begin{equation}
|E_\mathrm{sp}\rangle =\frac{1}{\sqrt{2}}\left[  \left| \begin{matrix}
\bullet & \circ \\
\circ & \bullet 
\end{matrix} \right\rangle
-
\left| \begin{matrix}
\circ & \bullet  \\
\bullet & \circ 
\end{matrix}  \right\rangle
\right], 
\end{equation}
where $|\circ\rangle$ and $|\bullet\rangle$ represent the ground and excited states of a qubit, respectively. Then, when such tetramers are connected with a weak coupling $J_\mathrm{e}$, a pair of special Fock product states, $|{\mathbf s}_0^S\rangle$ and $|{\mathbf s}_0^{S\prime}\rangle$, partially inherit properties from the single tetramer. Thus, for a half-filling  $4\times 6$ system, a scarred initial product state is given by  
\begin{equation}
    |{\mathbf s}_0^S\rangle = 
    \begin{matrix}
    \lceil\bullet  & \circ\rceil     & \lceil\circ     & \bullet\rceil  &  \lceil\bullet   & \circ\rceil  \\
    \lfloor\circ   & \bullet \rfloor & \lfloor\bullet  &  \circ\rfloor &  \lfloor\circ  & \bullet \rfloor \\
    \lceil\circ & \bullet\rceil  & \lceil\bullet   & \circ\rceil     &  \lceil\circ    & \bullet\rceil   \\
    \lfloor\bullet  & \circ \rfloor & \lfloor\circ    &  \bullet\rfloor & \lfloor\bullet & \circ \rfloor  \\
  \end{matrix},
\end{equation}
where the block marks a tetramer. Each tetramer has a $\pi$-phase difference from its adjacent ones.
As a system has the particle-hole symmetry, the Fock product state $|{\mathbf s}_0^{S\prime}\rangle$ has the same properties with the inverse state for each qubit. 

\begin{figure}
  \includegraphics[width=1.0\columnwidth]{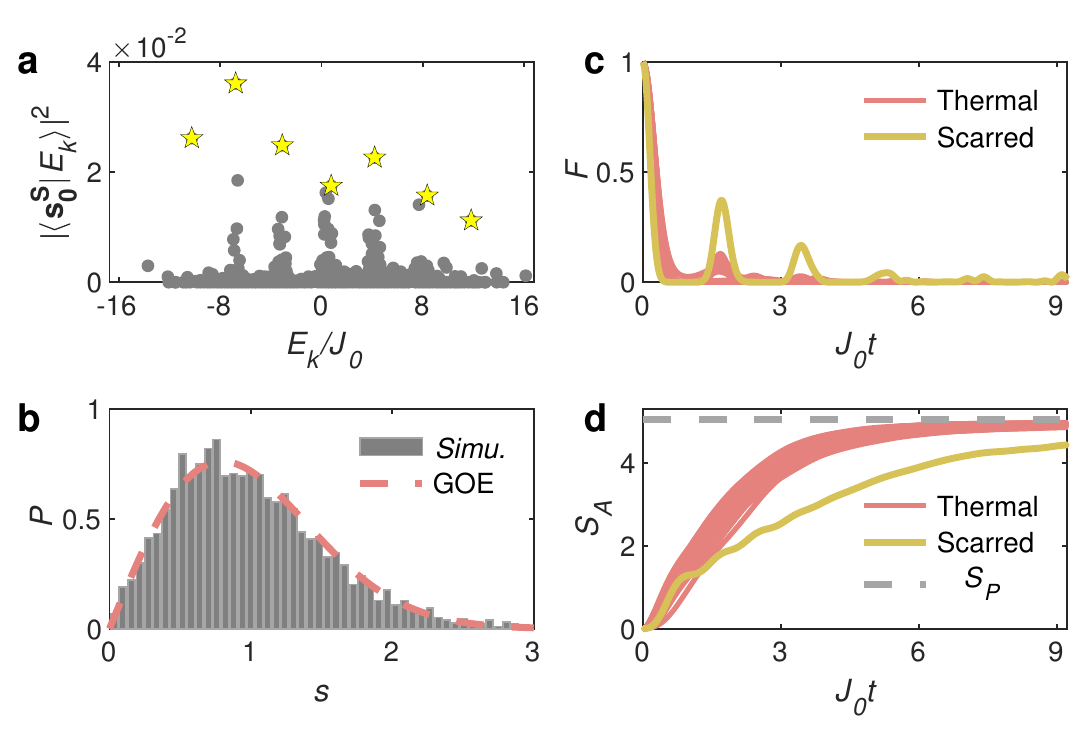}
      \caption{
      {\bf Properties of 2D scar.} 
      Panel~{\bf a} shows the overlap between the scarred state $|{\mathbf s}_0^S\rangle$ and all eigenstates as a function of eigenenergy $E_k$. Yellow pentagrams mark the towers' peaks, representing a set of special scarred eigenstates. 
      Panel {\bf b} shows the level spacing distribution in a particle-hole symmetry sector. Compared to the GOE curve, this simulation result implies the absence of integrability. Panels {\bf c} and {\bf d} show the dynamics of fidelity and bipartite EE, respectively, for different initial product states. Red curves represent $20$ randomly chosen thermal states, while yellow one represents the scarred state. Gray dashed line in {\bf d} is the Page value of bipartite EE, $S_P$. Both {\bf c} and {\bf d} show that the scarred state has a slow thermalizing behavior. All simulation results are based on a disorder-free 2D SSH model in  $4 \times 4$ lattice with couplings $J_\mathrm{o}/2\pi = 2J_\mathrm{e}/2\pi=-6$~MHz and $g_{\mathrm{x}}/2\pi = 0.9$ MHz.
}
  \label{Figure:2DScar_depict}
\end{figure}

Product states $|{\mathbf s}_0^S\rangle$ and $|{\mathbf s}_0^{S\prime}\rangle$ have remarkable overlap with a set of special eigenstates, which are equally spaced in the energy spectrum. The numerical simulation of energy level-spacing distribution for $4\times4$ lattice agrees with the Wigner-Dyson type, as shown in Fig.~\ref{Figure:2DScar_depict}. If we prepare the initial state as $|{\mathbf s}_0^S\rangle$ or $|{\mathbf s}_0^{S\prime}\rangle$, their fidelity dynamics show revival behaviors, and the dynamics of bipartite entanglement entropy (EE) also shows a slower growth. On the contrary, other product states show a thermalizing behavior, implied by their fidelity and bipartite EE dynamics, in which their initial information ergodically disperses in the whole Hilbert space. 

\section{Local observables in real space}
\begin{figure}
  \includegraphics[width=1\columnwidth]{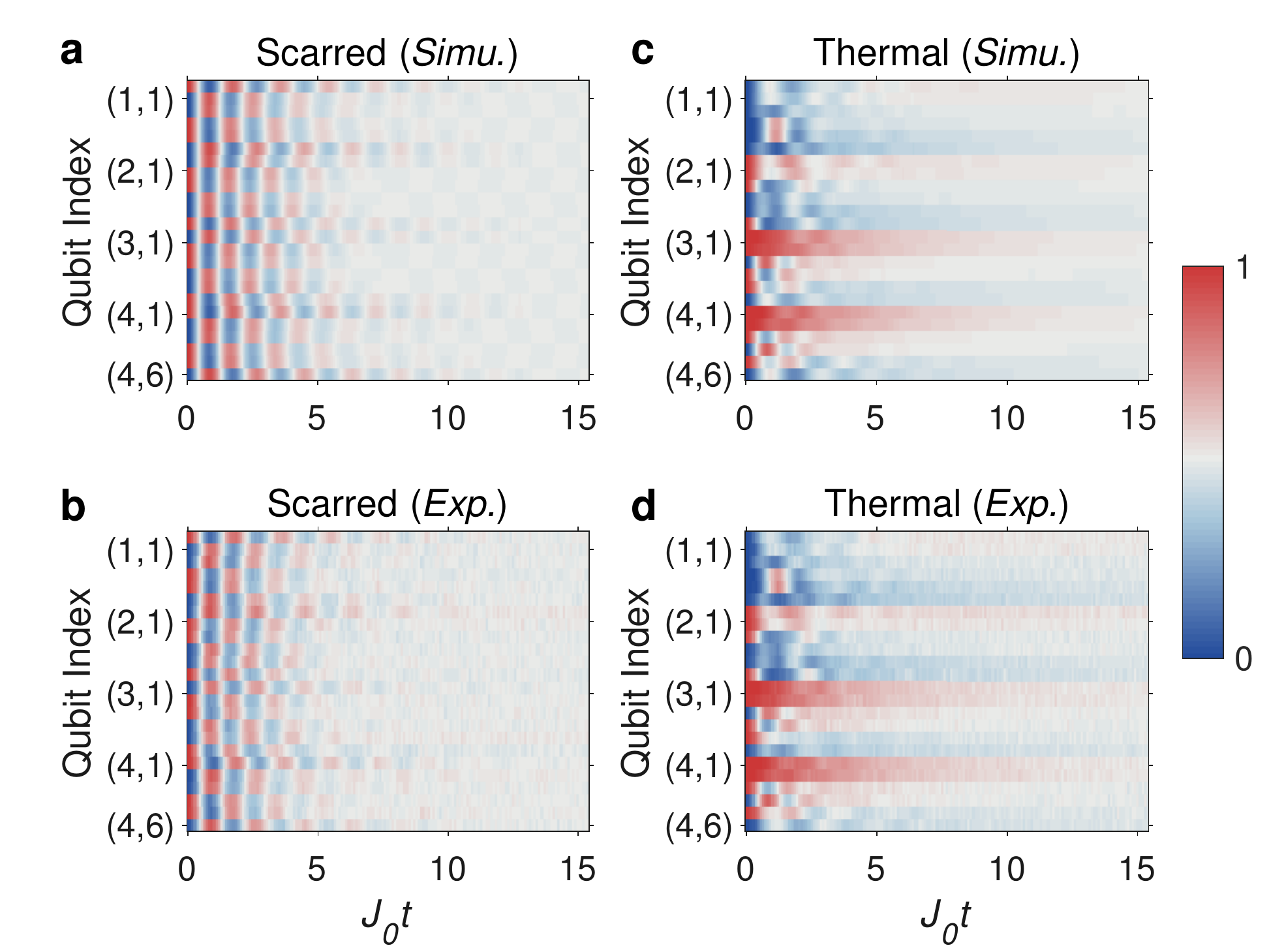}
   % \vspace{-0.6cm}
   \caption{
   {\bf Population dynamics without disorder.} 
   Panels {\bf a} and {\bf b} show the numerical and experimental population dynamics for the scarred state $|{\mathbf s}_0^S\rangle$, respectively. Panels {\bf c} and {\bf d} show similar results for the thermal state  $|{\mathbf s}_0^T\rangle$.
   }
   \label{Figure:popu_scar_thermal}
\end{figure}

\begin{figure}[h]
  \includegraphics[width=1\columnwidth]{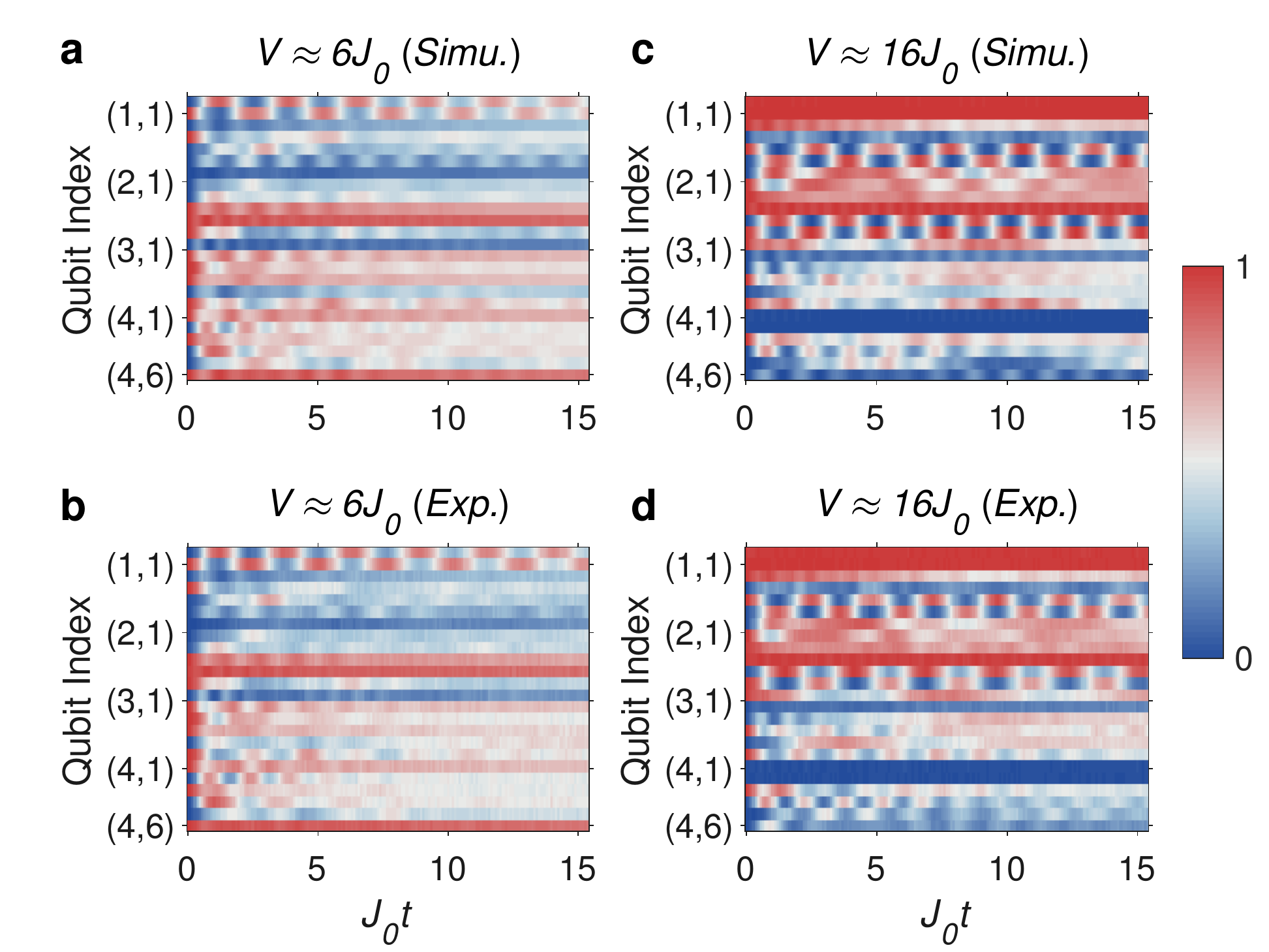}
   % \vspace{-0.6cm}
   \caption{
   {\bf Population dynamics with disorder.} 
    Panels {\bf a} and {\bf b} show the numerical and experimental population dynamics for $V \approx 6J_{0}$, respectively.
   Panels {\bf c} and {\bf d} are the similar results for a much larger disorder strength $V \approx 16J_{0}$. 
    } 
   \label{Figure:MBL}
\end{figure}

\begin{figure}[h]
  \includegraphics[width=1\columnwidth]{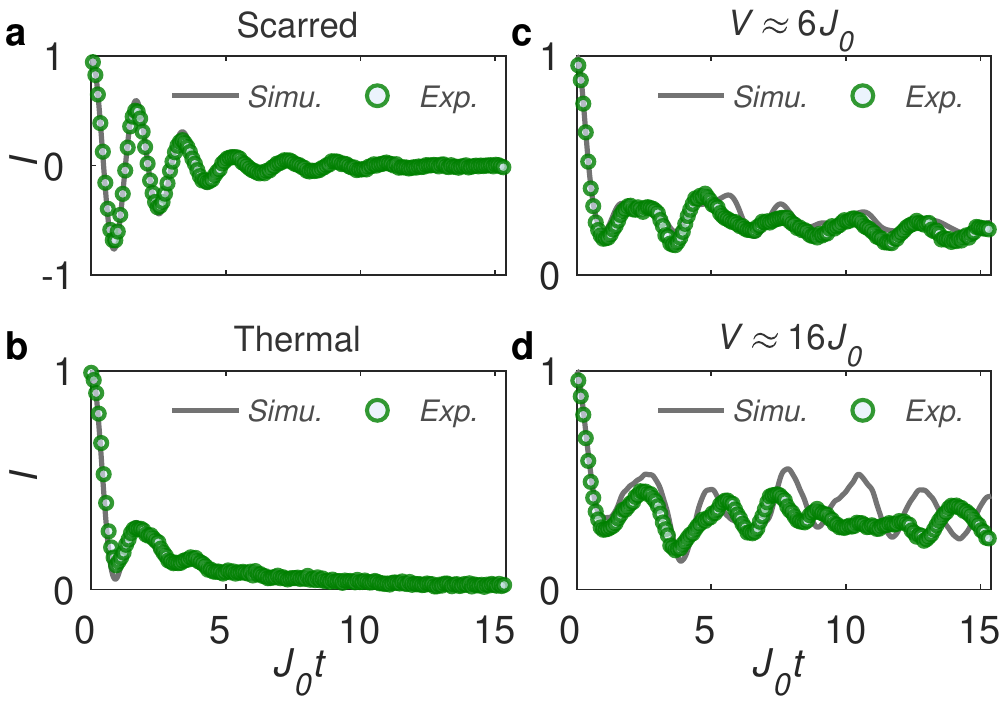}
   % \vspace{-0.6cm}
   \caption{
   {\bf Imbalance dynamics.} 
   Panels {\bf a} and {\bf b} show imbalance $I(t)$
   for the scarred state $|{\mathbf s}_0^S\rangle$ and the thermal state $|{\mathbf s}_0^T\rangle$ at $V=0$, respectively. Panels {\bf c} and {\bf d} show the $I(t)$ in disordered systems with $V\approx 6J_{0}$ and $ 16J_{0}$, respectively.  Green circle markers represent the experiment data, and gray lines represent the numerical data.
   %\ly{y axis is betten in log scale. Now there is no big difference for panels b-d.}
   }
   \label{Figure:imb}
\end{figure}

In general, measuring local observables in real space is the conventional way to study the quantum many-body dynamics experimentally. The population dynamics $p_m(t)$ for each qubit in real space help us to identify many-body states. As shown in Fig.~\ref{Figure:popu_scar_thermal} and Fig.~\ref{Figure:MBL}, the population dynamics for both experimental data and numerical data are in good agreement with each other. For the scarred initial state $|{\mathbf s}_0^S\rangle$, the populations show remarkable oscillating patterns (Fig.~\ref{Figure:popu_scar_thermal}{\bf a}, {\bf b}). In contrast, populations for all the qubits rapidly approach to a stable value of 0.5 for the thermal initial state  $|{\mathbf s}_0^T\rangle$ (Fig.~\ref{Figure:popu_scar_thermal}{\bf c}, {\bf d}). In the presence of disorders, the dynamics can preserve the initial populations for a long time, as shown in Fig.~\ref{Figure:MBL}. In addition to the population dynamics, a quantity known as the generalized imbalance $I(t)$~\cite{Guo2021np} is also commonly used to describe the preservation of initial local information in real space. Its definition is given by
\begin{equation}
  I(t)=\frac{1}{L}\sum_{m=1}^L\langle \sigma_m^z(t)\rangle\langle\sigma_m^z(0)\rangle,%=\frac{1}{L}\sum_{i=1}^L(-1)^{2n_i(0)-1}\langle \sigma_i^z(t)\rangle,
\end{equation}
where $\langle\sigma_m^z(t)\rangle=2p_m(t)-1$. We note that initial states are prepared as Fock states in the half-filling condition and $I(t=0)=1$. As expected, the scarred dynamics exhibits strong oscillations and remain for a considerable period of time~\cite{articlescar} (Fig.~\ref{Figure:imb}{\bf a}), while the thermalizing dynamics quickly drop to zero~(Fig.~\ref{Figure:imb}{\bf b}). For the disordered system, the imbalance tends to stabilize at a finite value much larger than zero~(Fig.~\ref{Figure:imb}{\bf c}  and {\bf d}).

\section{Few-body entanglement entropy}
For a quantum many-body system with a system size of $L$, it can be decomposed into two subsystems A and B in real space, with subsystem size ${l_A}$ and ${l_B=L-l_A}$. The reduced density matrix of subsystem A is given by $\rho_{A}=\mathrm{tr}_{B}(\rho)$, where $\rho$ is the density matrix for the whole system. Then, the von Neumann entanglement entropy is given by
\begin{equation}
S_{A}=-\mathrm{tr}\left(\rho_{A} \log \rho_{A}\right),
\end{equation}
which depicts the entanglement complexity between two subsystems. 

\begin{figure}
  \includegraphics[width=0.9\columnwidth]{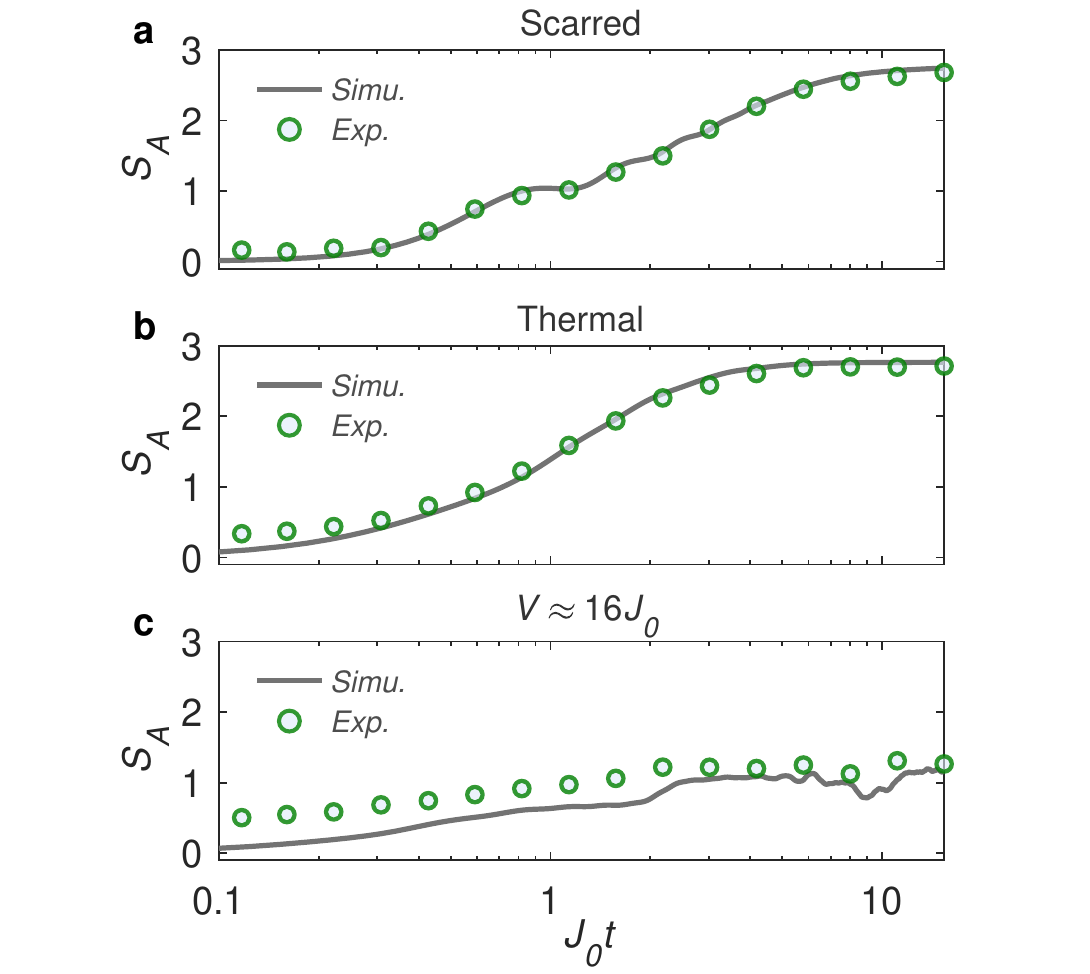}
   % \vspace{-0.6cm}
      \caption{
      {\bf Time evolution of four-qubit entanglement entropy.} 
      Four-qubit EE as a function of evolution time in log scale are shown in panel {\bf a} for the scarred state $|{\mathbf s}_0^S\rangle$ ($V=0$),  panel {\bf b} for the thermal state $|{\mathbf s}_0^T\rangle$ ($V=0$), and panel {\bf c} for the nonergodic dynamics of $V\approx 16J_{0}$. Gray lines represent the simulation results, while green circles represent the experiment data.
      }
   \label{Figure:EE}
\end{figure}

The experimentally measured four-qubit ($l_A=4$) EE are shown in Fig.~\ref{Figure:EE} for the scarred, thermal, and localized states. For the scarred state, the growth of EE is slower than the thermalizing dynamics but eventually reaches the similar thermal value. In the presence of disorder ($V\approx 16J_0$), the EE grows slowly  and tends to saturate to a value much smaller than thermal value,  indicating the many-body localization in our finite-sized system. 

\section{Limitation of few-body observables}
\begin{figure}[b]
  \includegraphics[width=1\columnwidth]{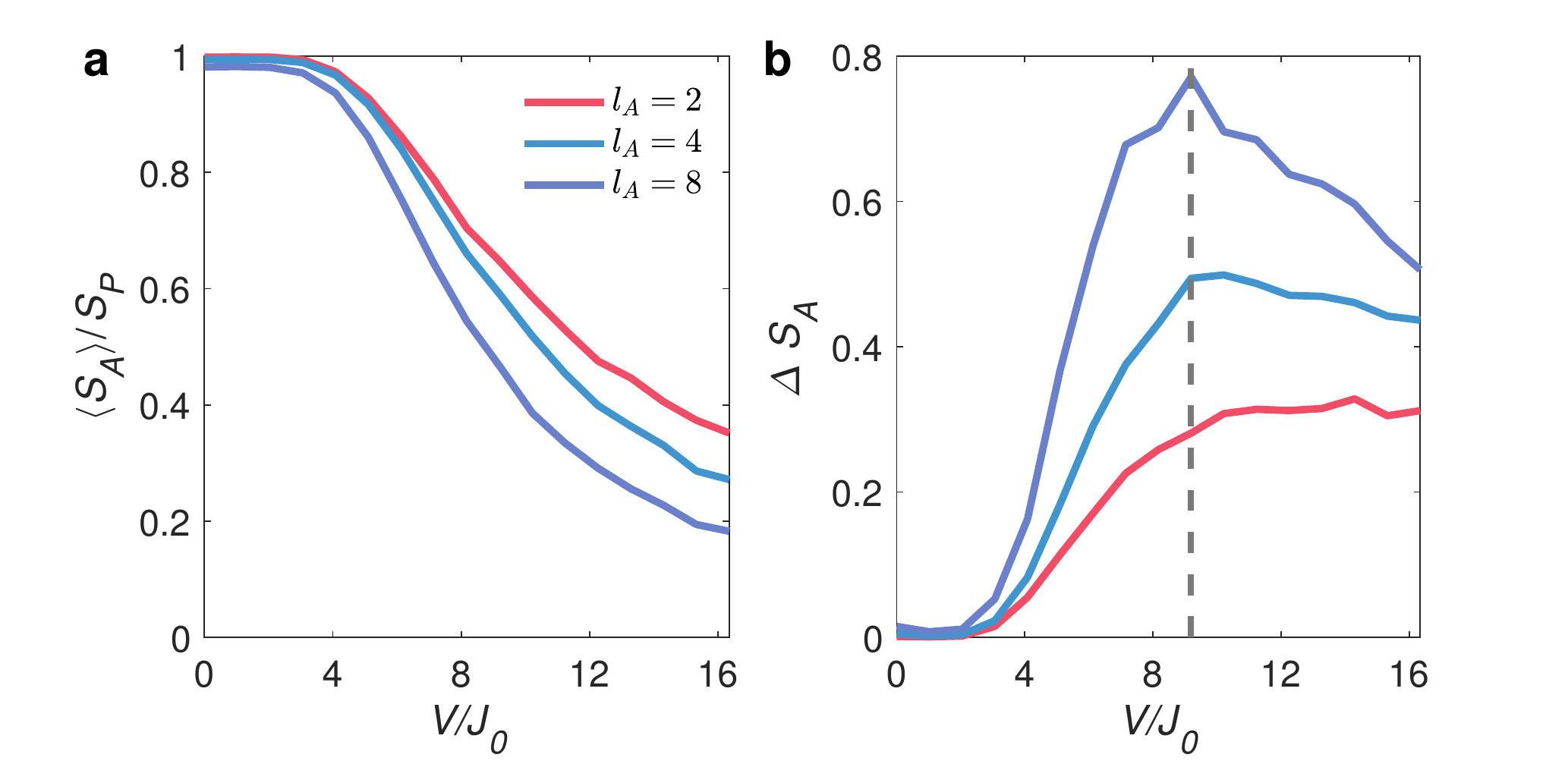}
   % \vspace{-0.6cm}
      \caption{{\bf Numerical entanglement entropy for different subsystem sizes.}
      Panel {\bf a} and {\bf b} show the disorder-averaged ($k=800$) subsystem entanglement entropy $\langle S_A\rangle$ and its standard deviation $\Delta S_A$ for $4\times4$ SSH model ($J_\mathrm{o}/2\pi=2J_\mathrm{e}/2\pi=-6$~MHz, $g_{\mathrm x}/2\pi=0.9$~MHz) at $t=1000$~ns, respectively. $l_A$ represents the site number of the subsystem $A$. Compared with few-body results~($l_A= $2, 4), the bipartite EE ($l_A=L/2=8$) is much more efficient in depicting the critical behavior of the MBL transition. The estimated critical disorder with bipartite EE is marked by the gray dashed line. 
      %\ly{y-axis label, $S_\mathrm{Page}$, not $S_{page}$}
      }
   \label{Figure:SA_size}
\end{figure}

Even though few-body EE is experimentally measurable, compared to global quantities~(e.g., bipartite EE), it may fail to characterize the entanglement information of the whole system precisely~\cite{Hamazaki_2018}. Fig.~\ref{Figure:SA_size} displays numerical results of EE with different subsystem size $l_A$ for $4\times4$ square lattice. Obviously, the bipartite ($l_A=L/2$) EE shows the sharpest transition peak from thermalization to localization. On the contrary, the standard deviation of two-qubit EE totally misses the transition point, and the four-qubit case shows a broad range of indistinctive enhancement of $\Delta S$, leading to an inaccurate estimation of critical disorder strength. Despite the reliability of bipartite EE, it is exponentially hard for numerics and experiments with the growth of system size, which makes its scaling analysis impossible for large systems.

\begin{figure*}
  \includegraphics[width=2.0\columnwidth]{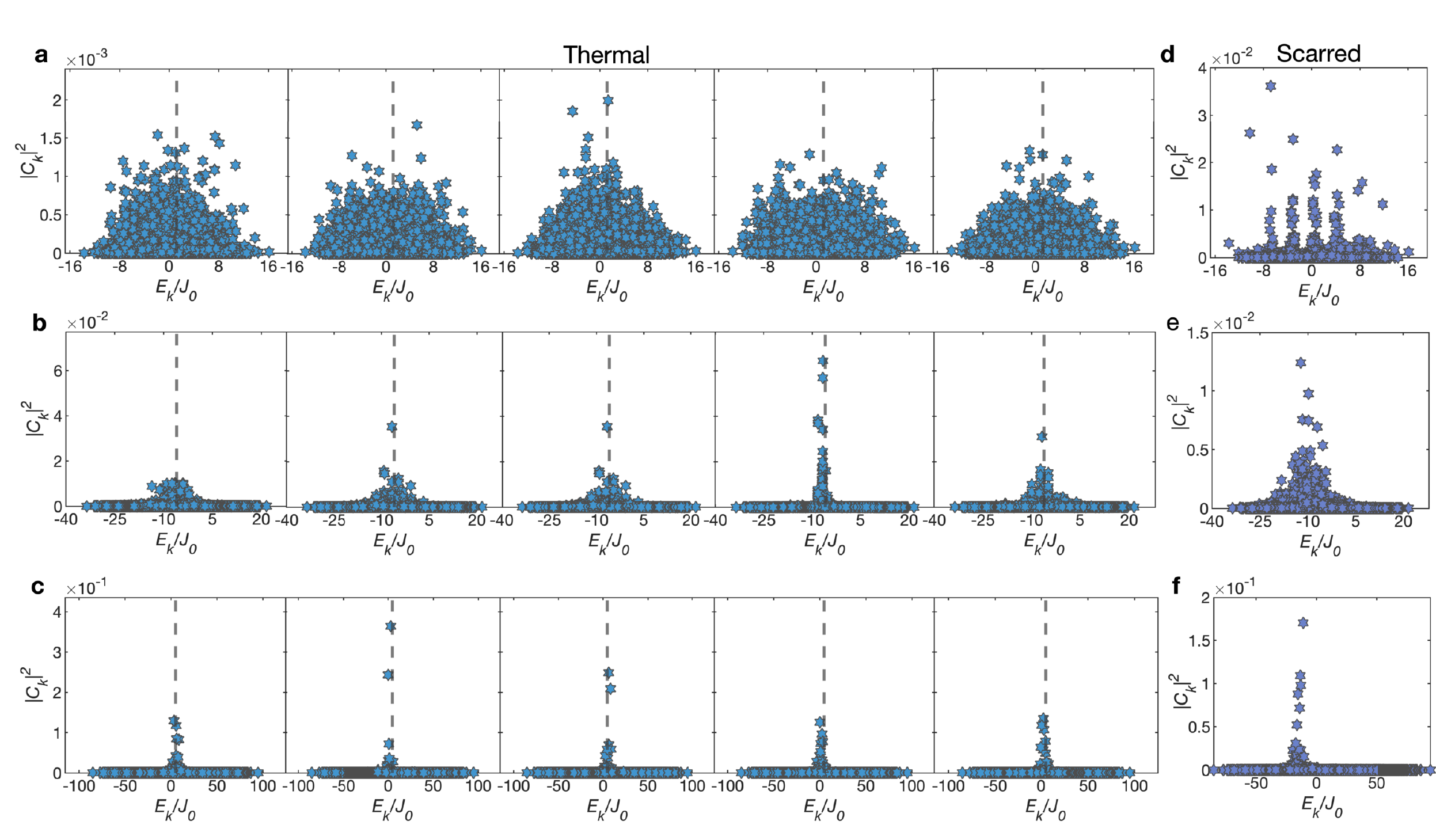}
   % \vspace{-0.6cm}
   \caption{
   {\bf Overlaps between initial states and eigenstates.} 
    Panels {\bf a}, {\bf b}, and {\bf c} show the overlaps $|C_k|^2 = |\langle {\mathbf s}_0|E_k\rangle|^2$ between five randomly selected initial states and eigenstates in $4\times4$ 2D SSH system ($J_\mathrm{o}/2\pi=2J_\mathrm{e}/2\pi=-6$~MHz, $g_{\mathrm x}/2\pi=0.9$~MHz), with disorder strength $V = 0$, $V\approx 6J_{0}$, and $V\approx 16J_{0}$, respectively. Each state is randomly selected in the energy window $[E_\mathrm{mid}-\Delta E/100, E_\mathrm{mid}+\Delta E/100]$. The gray dashed line indicates the center of the energy spectrum. 
    Panel {\bf d} shows the overlap spectrum for the scarred state $|{\mathbf s}_0^S\rangle$ at $V=0$, where multi-tower feature with higher overlap on some special eigenstates appears. Panels {\bf e} and {\bf f} are the similar results at $V\approx 6J_{0}$ and $V\approx 16J_{0}$, respectively. With the growing disorder strength, the multi-tower feature fades and keeps only one peak near its total energy.}
   \label{Figure:overlap}
\end{figure*}
\section{Selection of initial states}\label{sec:initial_state}
Fock product states $|\mathbf{s}\rangle$ are friendly to prepare for experiments. However, after long-time evolution, not all of them have similar properties to the eigenstates of the middle spectrum due to the possible existence of many-body mobility edges~\cite{Luitz2015PRB, Guo2021np}. In our work, we choose the Fock product states in the center of the energy spectrum. The energy of each Fock state can be obtained easily by $E/\hbar=\sum_{m, Q_m=|1\rangle}V_m$. Since the energy spectrum of the Hamiltonian for larger systems is approximately Gaussian distributed, we can approximate the energy in the middle by numerically averaging the maximum and minimum eigenvalues ($E_{\mathrm{mid}}=(E_{\mathrm{max}}+E_{\mathrm{min}})/2$). Thus, we randomly select one Fock state with energy $E$ in the range of $[E_{\mathrm{mid}}-\Delta E/100, E_{\mathrm{mid}}+\Delta E/100]$ for each realization, where $\Delta E=E_\mathrm{max}-E_\mathrm{min}$ is the bandwidth of energy spectrum.

To examine the effectiveness of the selection strategy, as shown in Figs.~\ref{Figure:overlap}{\bf a}, {\bf b}, and {\bf c}, we plot the overlaps between five randomly selected initial states and all eigenstates for  $4\times 4$ 2D SSH model for different disorders. We find that the middle eigenstates have much larger overlaps with our initial states than the marginal ones. Furthermore, as the disorder strength increases, the distribution of overlap over energy eigenstates becomes narrower. Figs.~\ref{Figure:overlap}{\bf d}, {\bf e}, and {\bf f} display the overlap for the scarred initial state $|{\mathbf s}_0^S\rangle$. For $V=0$, it shows a multi-peak behavior (Fig.~\ref{Figure:overlap}{\bf d}). In the presence of large disorder $V\approx16J_0$, all these peaks shrink to a sharp one ~(Fig.~\ref{Figure:overlap}{\bf f}). Its position deviates a little from the center due to the change of total energy caused by disorders.  

\section{Derivation of ergodic wave packet}
Our model is a hard-core Bose-Hubbard model, and the total particle number is conserved during the time evolution. In the half-filling condition, $\Pi(d)$ vanishes for the odd Hamming distance $D(\mathbf{s},\mathbf{s}_0)$, since a photon hopping from one site to another sit leads to a change of two for $D(\mathbf{s},\mathbf{s}_0)$. Thus, values of Hamming distance are a set of even integers as $d=0,2,4,\cdots,L$ for our system.

For the ideal ergodic wavefunction, its overlap with each Fock-space site is the same. Thus, we have a hyper-geometric-like distribution as
\begin{equation}
\Pi^\mathrm{Erg.}(d)
=\frac{C_{L/2}^{d/2}C_{L/2}^{L/2-d/2}}{C_{L}^{L/2}}.
\end{equation}
Then,  the displacement is  given by
\begin{equation}
    \overline{d} =\sum_d\Pi(d)\cdot d=\frac{1}{2}\sum_d\frac{C_{L/2}^{d/2}C_{L/2}^{L/2-d/2}}{C_{L-1}^{L/2-1}}\cdot d = \frac{L}{2}.
\end{equation}
Variance is defined as $\Delta d^2=\overline{d^2}-\overline{d}^2$ with $\overline{d^2}=\overline{d(d-1)}+\overline{d}$. We have
\begin{equation*}
  \begin{split}
    \overline{d(d-1)}&=\sum_{d}\frac{C_{L/2}^{d/2}C_{L/2}^{L/2-d/2}}{C_{L}^{L/2}}d(d-1)\\
    &=\frac{L}{4}\frac{L^2-2L+2}{L-1},
  \end{split}
\end{equation*}
thus, 
\begin{equation*}
    \Delta d^2=\overline{d(d-1)}+\overline{d}-\overline{d}^2=\frac{L^2}{4(L-1)}.
\end{equation*}
In conclusion, for the ergodic wave packet, normalized displacement $\mathrm{x^{\rm Erg.}}=\overline{d}/L=0.5$ and normalized width $\Delta \mathrm{x}^{{\rm Erg.}}=\frac{\sqrt{L-1}}{L}\sqrt{\Delta d^2}=0.5$. Moreover, the system fluctuation $\sigma$, as it has been introduced in the main text, would be $\sigma=\frac{\sqrt{L-1}}{L}\sqrt{\sum_{d}d^2\Pi(d)-\langle \mathrm{x}\rangle^2}=0.5$.

\section{Numerical simulation of $4\times4$ lattice}

\begin{figure}
    \includegraphics[width=1.0\columnwidth]{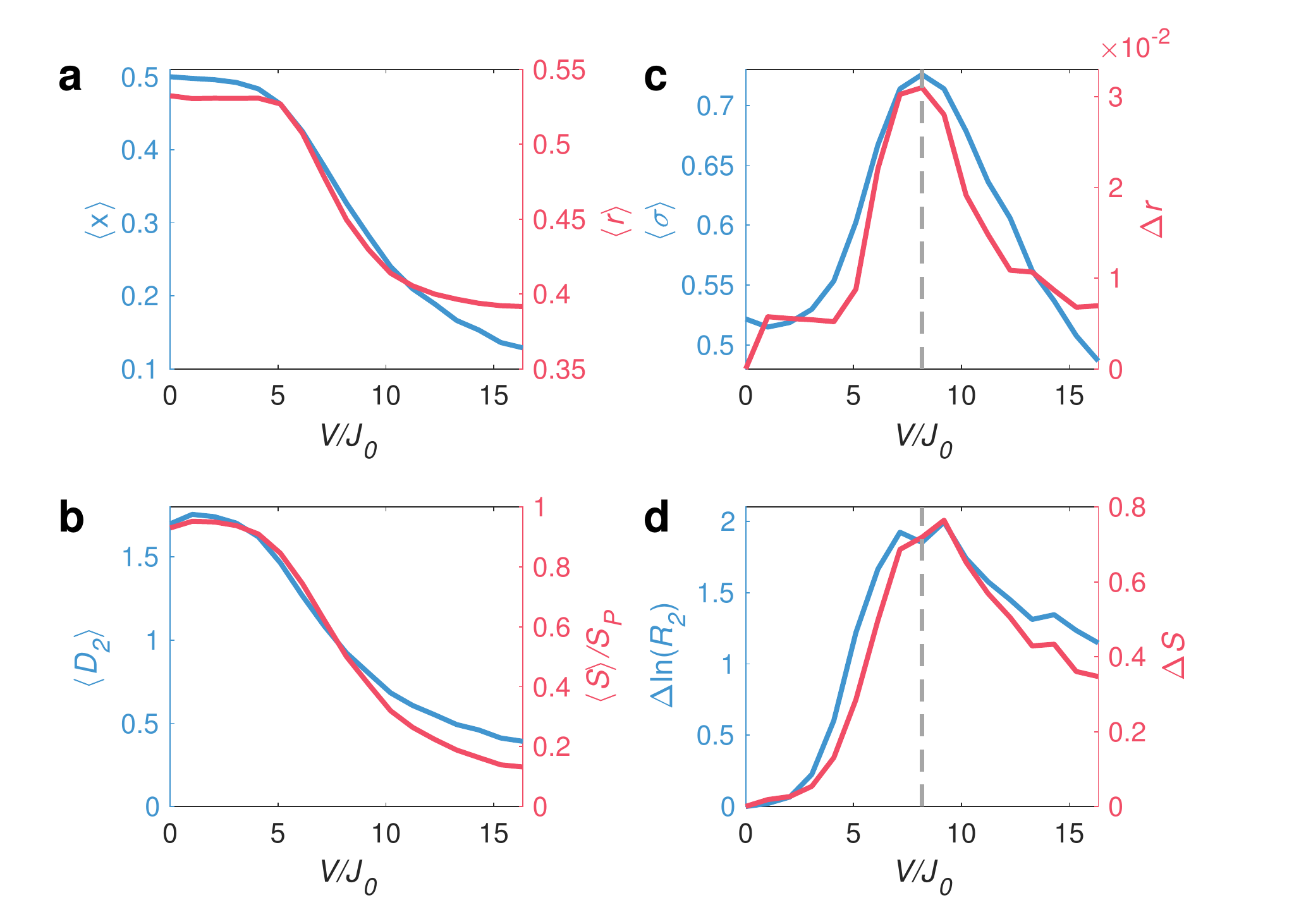}
    \caption{{\bf Numerical simulation of MBL indicators with eigenstate properties.} 
    {\bf a}. Disorder-averaged normalized displacement $\langle \mathrm{x} \rangle$ (blue) and level spacing ratios $\langle r \rangle$ (red) as a function of disorder strength $V$.
    {\bf b}. Disorder-averaged $2$-moment generalized fractal dimension $\langle D_2\rangle$ (blue) and rescaled bipartite EE $\langle S\rangle/S_P$ (red) as a function of disorder strength $V$.
    {\bf c}, Fluctuation of wave packets $\langle\sigma\rangle$ (blue) and standard deviation of level spacing ratios $\Delta r$ (red) as a function of disorder strength $V$. 
    {\bf d}, Standard deviation of bipartite EE $\Delta S$ (red) and standard deviation of 2-moment IPR $\Delta \mathrm{ln} \mathrm{(R_2)}$ (blue) as a function of disorder strength $V$. The gray dashed lines in {\bf c} and {\bf d} indicate the critical point of the finite-size MBL transition for $4\times4$ lattice. The model for numerics is the $4\times4$ SSH lattice with $J_\mathrm{o}/2\pi=2J_\mathrm{e}/2\pi = -6~\mathrm{MHz}$ and $g_{\mathrm{x}}/2\pi=0.9$~MHz. For each $V$, we randomly choose $k=800$ disorder realizations and randomly select ten eigenstates in the middle energy spectrum for each of them.
    } 
    \label{FigureS:IPR}
\end{figure}
In this section, we numerically compare our Fock-space wave packet approach with the conventional observables, such as level statistics, inverse participation ratio, and entanglement entropy on a $4\times4$ 2D SSH lattice. 

In the random matrix theory, the statistics of level spacing $s_i$ between two nearest eigenenergies $E_i$ and $E_{i+1}$ is a standard way to differentiate ergodic and nonergodic systems. For a quantum chaotic system, its distribution shows a Gauss orthogonal ensemble (GOE) type, $P_{\mathrm{GOE}}=\frac{\pi}{2} s \mathrm{e}^{-(\pi / 4) s^{2}}~$\cite{GUHR1998189}, whereas a Poisson distribution, $P_{\mathrm{Poisson}}(s)=e^{-s}$, is expected when a quantum many-body system is localized. Different kinds of level spacing statistics can be characterized by the level spacing ratio $r_{i}=\min \left({s_{i+1}}/{s_{i}}, {s_{i}}/{s_{i+1}}\right)$. The mean value $\langle r \rangle$ at central energy spectrum regime, levels off $(2 \mathrm{ln}2 -1)\approx0.386$ or $(4-2 \sqrt{3}) \approx 0.536$ for an integrable or a GOE distribution, respectively~\cite{IZRAILEV1990299}. Figure~\ref{FigureS:IPR}{\bf a} shows that $\langle r\rangle$ is equal to about $0.53$ for weak disorder regime ($V<4J_{0}$) and drops to  the Poisson distribution value $0.39$ for the strong disorder regime near $V\approx 16J_{0}$. For comparison, the disorder-averaged normalized displacement $\langle \mathrm{x} \rangle$ is also plotted in Fig.~\ref{FigureS:IPR}{\bf a}.  

In Fock space, the multifractal analysis of inverse participation ratio (IPR) is a popular way to characterize MBL transition. Here, we use $2$-moment generalized fractal dimension, which is given by~\cite{Evers_2008}
\begin{equation}
    {D}_{2}=-\ln R_{2} / \ln \mathcal{N},
\end{equation}
where the inverse participation ratio $R_{2} = \mathrm{IPR}_2 =\sum_{\alpha}\left|\psi_{\alpha}\right|^{4}$. $\{|\alpha\rangle\}$ is a set of orthonormal basis of an $\mathcal{N}$-dimension Hilbert space and $|\psi\rangle=\sum_{\alpha} \psi_{\alpha}|\alpha\rangle$. As shown in Fig.~\ref{FigureS:IPR}{\bf b}, ${D}_{2}$ decays monotonically as disorder strength grows and the bipartite EE also shows similar behaviors. 

To extract the critical disorder, the fluctuations, such as  $\langle\sigma\rangle$, $\Delta r$, $\Delta \mathrm{ln}(R_2)$, and $\Delta S$ are displayed in Figs.~\ref{FigureS:IPR}{\bf c} or {\bf d}. In spite of the slight difference, all of them indicate almost the same critical disorder near  $\sim8J_0$.

\section{Error mitigation}\label{sec:error_mig}

\begin{figure*}
    \includegraphics[width=2\columnwidth]{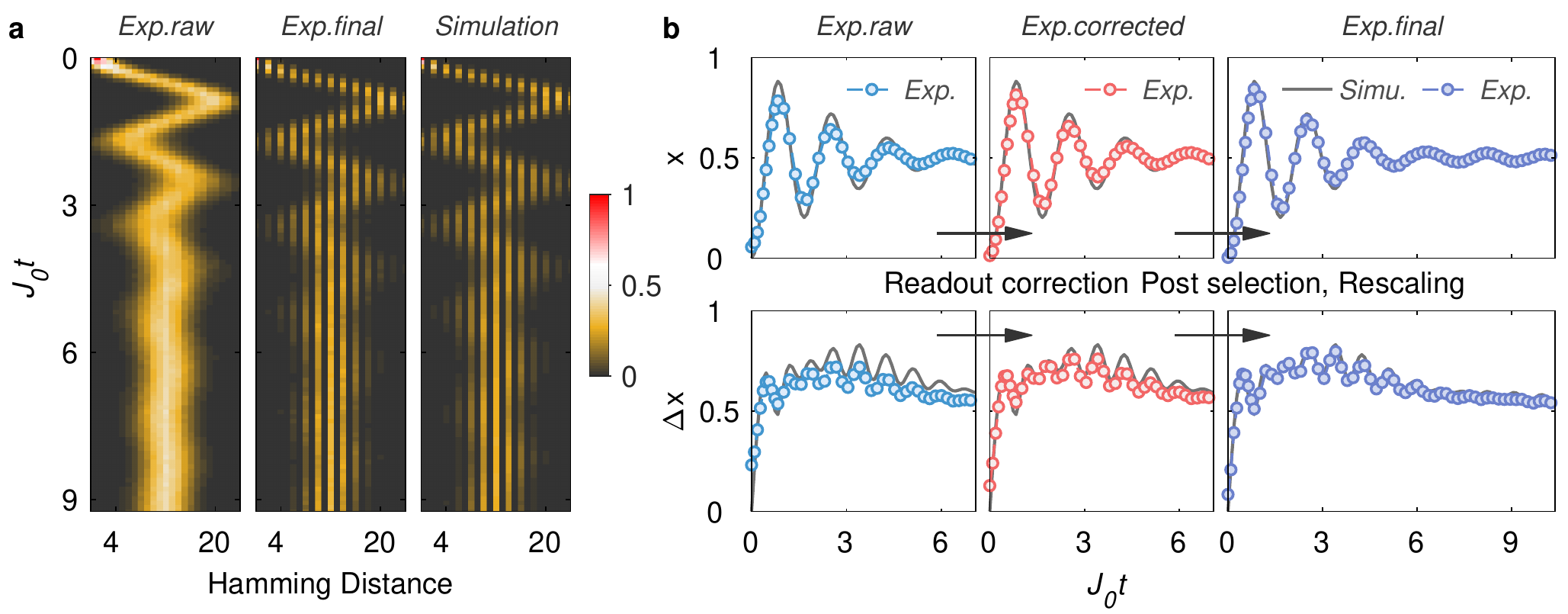}
    \caption{
    {\bf Effects of error mitigation.}
    {\bf a}. Experimentally measured wave packet dynamics are shown in the left panel for the results extracted with raw data, and in the middle panel for the results using error mitigation. For comparison, the numerical results are shown in the right panel, which almost perfectly match the error-mitigated experimental results. {\bf b.} Effects of readout correction and post-selection for normalized displacement $\mathrm{x}$ and normalized displacement $\Delta\mathrm{x}$. The results with readout correction are shown in the middle column, based on which the post-selection is further applied and the results are shown in the right column. The experimental and numerical data are obtained on the half-filling $4\times 6$ SSH model in the main text. 
    }
    \label{fig:error}
\end{figure*}

To mitigate state preparation and measurement (SPAM) errors,  and qubit energy relaxation ($T_1$) errors, our error mitigation scheme includes readout correction and post-selection in the data processing~\cite{Google2020Arxiv,Guo2021np}. Here, we illustrate  detailed information about them. 

 As shown in Table~\ref{T1}, we list  readout fidelities $\{F_{0, (m,n)}, F_{1,(m,n)}\}$ for each qubit. Thus, the readout correction matrix for qubit $Q_{(m,n)}$ is given by
\begin{equation}
S_{(m,n)}=\left(
\begin{array}{cc}
F_{0,{(m,n)}} & 1-F_{1,{(m,n)}} \\
1-F_{0,{(m,n)}} & F_{1,{(m,n)}}
\end{array}
\right).
\end{equation}
Then, the corrected probability vector for the system is given by $\vec{P}_{\rm corr} =\bigotimes_{(m,n)} S_{(m,n)}^{-1}\cdot \vec{P}$, where $\vec{P}$ is the raw probability vector. Moreover, since our model conserves the total photon number during the time dynamics, we could mitigate T1 induced photon leakage errors  by only keeping the photon-number-conserved subspace of $P_{\rm corr}$.

Thanks to the coherent oscillating dynamics of the scarred state, we can clearly see effects of our error mitigation scheme step by step. In Fig.~\ref{fig:error}, we compare the numerical results with experimental results with and without using the error mitigation scheme. Obviously, both readout correction and post-selection remarkably improve the quality of experimental data. 

\begin{figure*}
    \includegraphics[width=2\columnwidth]{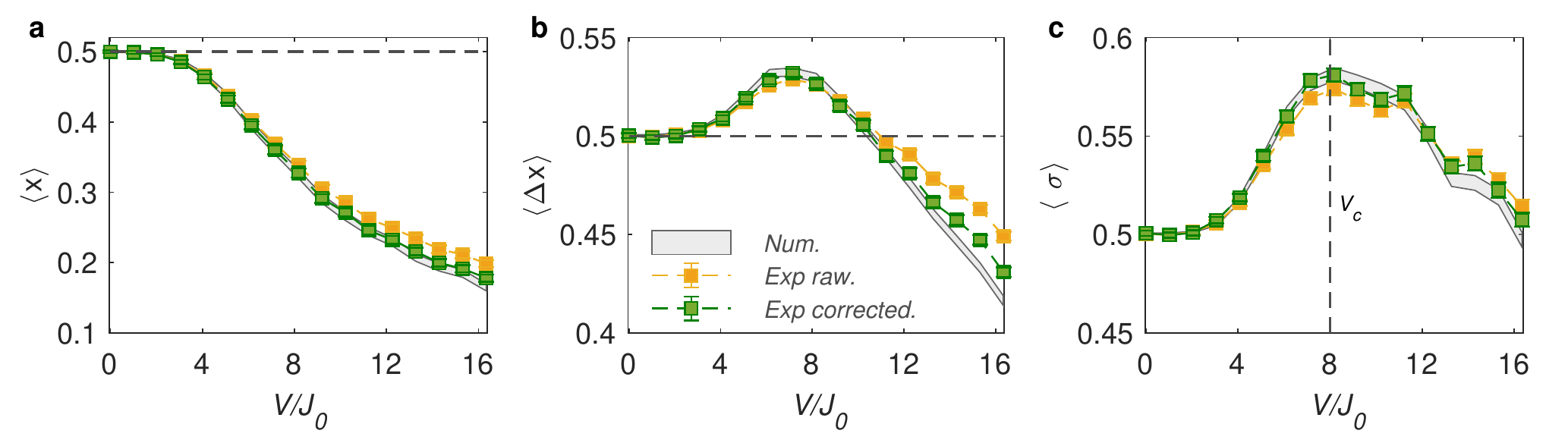}
    \caption{
    {\bf Robustness to errors.}
    {\bf a}, {\bf b}, and {\bf c} are long-time ($t=1000$~ns) disorder-averaged normalized displacement $\langle \mathrm{x}\rangle$, normalized width  $\langle\Delta \mathrm{x}\rangle$, and system fluctuation $\langle\sigma\rangle$ as a function of disorder strength $V$, respectively. Green squares are experimental data with error mitigation while yellow squares denote the raw data. The light gray areas represent numerical results. Despite some deviations, we can still clearly identify the critical regime and critical point. For each  $V$, we randomly choose $k=400$ disorder realizations.}
    \label{fig:error2}
\end{figure*}

In addition, we emphasize that the observation of critical behavior of MBL transition with  $\Delta \mathrm{x}$ and $\sigma$ are quite robust to SPAM errors and energy relaxation errors. It is a crucial point for a scalable protocol, since the readout correction scheme above is inaccessible for large systems. In Fig.~\ref{fig:error2}, we compare the raw data and the corrected data. The differences mainly appear in the strong disorder regime, while the critical regime and the critical point almost do not change . 

\section{Effect of decoherence}
Decoherence is inevitable for current experimental platforms, which is caused by the small couplings between the system and the surrounding environments. Decoherence has two effects on qubits: energy relaxation and dephasing. The energy relaxation is characterized by the energy relaxation time $T_{1}$, as listed in Table.~\ref{T1}. $T_{1}$~($\sim120~\mu$s) in our device is far larger than the maximum experimental time~(1~$\mu$s) and we use the post-selection method to mitigate its effect (see Sec.~\ref{sec:error_mig}). 

The effect of dephasing is much more complicated, especially for the interacting many-body system with all qubits actively coupled. Here, we demonstrate that the true dephasing times describing the coupled qubits are much longer than that measured by Ramsey experiments ($T_2^*$), whose timescales~($\sim 20~\mu$s) are similar to spin echo dephasing times $T^{SE}_2$. 

\begin{figure}
  \includegraphics[width=0.9\columnwidth]{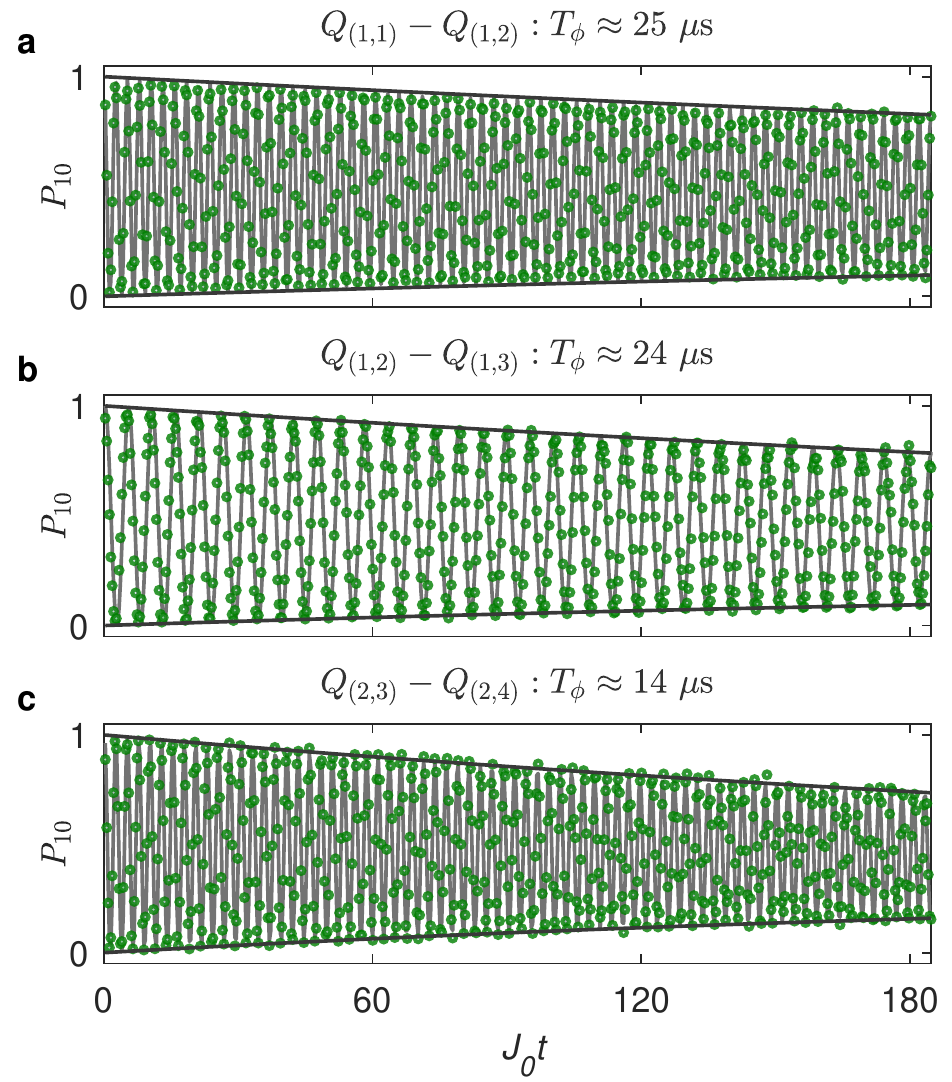}
   \caption{{\bf Effective dephasing time $ T_{\phi}$ for coupled qubit pairs.}
  $P_{10}$ dynamics of qubit pairs are shown in {\bf a} for $Q_{(1,1)}$-$Q_{(1,2)}$,  {\bf b} for $Q_{(1,2)}$-$Q_{(1,3)}$, and  {\bf c} for $Q_{(2,3)}$-$Q_{(2,4)}$. Green circles denote the experimental data and gray lines are the corresponding fitting data using Eq.~\ref{eq:dephasingTime}, which give the effective dephasing time about $25~\mu$s, $24~\mu$s, and $14~\mu$s for them.
   }
   \label{Figure:qqSwap}
\end{figure}

We choose nearest-neighbor qubit pairs, i.e., $Q_{(1,1)}$-$Q_{(1,2)}$,  $Q_{(1,2)}$-$Q_{(1,3)}$, and  $Q_{(2,3)}$-$Q_{(2,4)}$ to measure two-qubit resonant swap dynamics for each pair at $4.57$~GHz. The coupling strength is set to $J_{\rm NN}/2\pi\approx -6$ MHz or $-3 $~MHz. The experimental results of swap dynamics are shown in Fig.~\ref{Figure:qqSwap}. We fit the probability $P_{10}$ with the following formula \cite{Guo2021np}  
\begin{equation}
	P_{10}(t) = \frac{1}{2} \cos(2 J_{\rm NN} t)e^{-\frac{t}{\overline T_{1}} - \frac{t}{T_{\phi}}} + \frac{1}{2}e^{-\frac{t}{\overline T_{1}}},
    \label{eq:dephasingTime}
\end{equation}
where $\bar T_{1}=2 T_{1,m}T_{1,m+1} / (T_{1,m}+T_{1,m+1})$ is the average energy relaxation lifetime for the qubit pair, and $T_{\phi}$ is the effective dephasing time during the interaction. In this way, the effective dephasing times for qubit pairs $Q_{(1,2)}$-$Q_{(1,3)}$, $Q_{(1,2)}$-$Q_{(1,3)}$ and $Q_{(2,3)}$-$Q_{(2,4)}$, are estimated with values of  25~$\mu$s, 24 $\mu$s, and 14 $\mu$s (see Fig.~\ref{Figure:qqSwap}), respectively, which are one order of magnitude larger than the maximum experimental time ($1~\mu$s). Thus, it is reasonable to approximately treat our system as a closed quantum system within the experimental timescale.

%The slightly mismatch between the experimental results and simulation for the strongly disordered system could be due to some other issues.

\section{Finite-time effect}
\begin{figure*}
    \includegraphics[width=2\columnwidth]{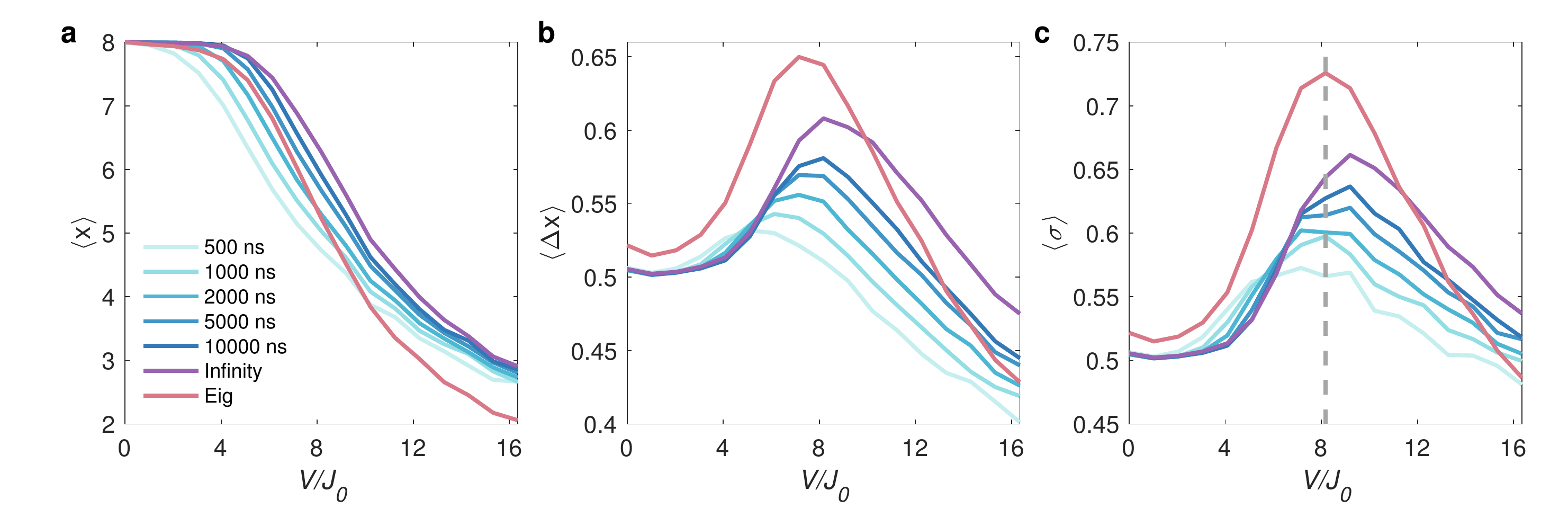}
    \caption{
    {\bf Finite-time effects.} 
    Numerical results at different timescales as well as eigenstate results are shown in panel {\bf a} for normalized displacement $\langle \mathrm{x}\rangle$, panel {\bf b} for normalized width  $\langle\Delta \mathrm{x}\rangle$, and panel {\bf c} for system fluctuation $\langle\sigma\rangle$. All of them are averaged over $k=800$ disorder realizations. Red curves represent the eigenstate results and the purple lines represent the results at infinite-time limit. The cyan-blue gradient curves show the time evolution of $500$~ns, $1000$~ns, $2000$~ns, $5000$~ns, and $10000$~ns. In panel {\bf c}, the dashed line labels the critical point estimated by eigenstate results. All the data are obtained on  $4\times4$ SSH model with $J_\mathrm{o}/2\pi=2J_\mathrm{e}/2\pi = -6~\mathrm{MHz}$ and $g_{\mathrm{x}}/2\pi=0.9$~MHz.
    } 
    \label{Figure:infinite_time}
\end{figure*}
In general, properties of eigenstates are the starting point for studying the MBL phase transition~\cite{Rigol2008Nature,Khemani2017PRX}. However, current experimental techniques only allow us to observe out-of-equilibrium quench dynamics within a finite timescale. The evolution time needs to be large enough in order to observe quantum thermalization and its breakdown. Therefore, it is necessary to demonstrate that the evolution time in our experiment ($1000$~ns) is sufficient to imply the nature of the system. We numerically investigate such an effect by varying the evolution time and compare the results in infinite-time limit obtained by eigenstates.

The relation between the eigenstates and time evolution of a quantum state is given by ($\hbar=1$)
\begin{equation}
    |\psi(t)\rangle=e^{-i \hat{H} t}|\psi(0)\rangle=\sum_k C_k e^{-i E_k t}|E_k\rangle,
\end{equation}
where $|\psi(t)\rangle$ is the time-dependent wavefunction. Therefore, an infinite-time operator $\langle\hat{O}\rangle_{t\rightarrow \infty}$ can be  written as
\begin{equation}
  \label{eq:infinite_time}
    \begin{aligned}
\langle \hat{O}\rangle_{t \rightarrow \infty} &\equiv \lim _{t \rightarrow \infty} \frac{1}{t} \int_0^t\langle\psi(\tau)|\hat{O}| \psi(\tau)\rangle d \tau \\
&\approx\sum_k|C_k|^2\langle E_k|\hat{O}| E_k\rangle .
    \end{aligned}
\end{equation}

According to Eq.~\ref{eq:infinite_time}, we can obtain the radial probability distribution at infinite time as
\begin{equation}
    \begin{aligned}
    \Pi(d)_{t\rightarrow\infty}&=\lim_{t\rightarrow\infty}\sum_{\mathbf{s} \in\left\{D\left(\mathbf{s}, \mathbf{s}_0\right)=d\right\}}\langle\psi\left(t\right)|\mathbf{s}\rangle\langle\mathbf{s}|\psi\left(t\right)\rangle\\
    &= \sum_k |C_k|^2\sum_{\mathbf{s} \in\left\{D\left(\mathbf{s}, \mathbf{s}_0\right)=d\right\}}|\langle E_k|\mathbf{s}\rangle|^2 ,
    \end{aligned}
\end{equation}
where $D\left(\mathbf{s}, \mathbf{s}_0\right)$ is the Hamming distance between state $|\mathbf{s}\rangle$ and $|\mathbf{s_0}\rangle$. Once we get the $\Pi(d)_{t\rightarrow\infty}$, we can calculate the wave packet displacement, width and fluctuation. 

The numerical results for disorder-averaged $\langle \mathrm{x}\rangle$, $\langle\Delta \mathrm{x}\rangle$, and $\langle \sigma\rangle$ at different times are shown in Fig.~\ref{Figure:infinite_time}, we find that the results at timescale in our experiment ($t=1000$ ns) have similar transition behaviors with eigenstates. We note that despite a higher peak of $\langle\Delta \mathrm{x}\rangle$ for eigenstate results, the peak position for identifying critical transition point is almost the same for both eigenstate results and finite-time results.

%\bibliographystyle{naturemag}
%\bibliography{mainRef.bib}

\begin{thebibliography}{10}
\expandafter\ifx\csname url\endcsname\relax
  \def\url#1{\texttt{#1}}\fi
\expandafter\ifx\csname urlprefix\endcsname\relax\def\urlprefix{URL }\fi
\providecommand{\bibinfo}[2]{#2}
\providecommand{\eprint}[2][]{\url{#2}}

\bibitem{Deutsch1991PRA}
\bibinfo{author}{Deutsch, J.~M.}
\newblock \bibinfo{title}{Quantum statistical mechanics in a closed system}.
\newblock \emph{\bibinfo{journal}{Phys. Rev. A}} \textbf{\bibinfo{volume}{43}},
  \bibinfo{pages}{2046--2049} (\bibinfo{year}{1991}).
%\newblock %\urlprefix\url{https://link.aps.org/doi/10.1103/PhysRevA.43.2046}.

\bibitem{Srednicki1994PRE}
\bibinfo{author}{Srednicki, M.}
\newblock \bibinfo{title}{Chaos and quantum thermalization}.
\newblock \emph{\bibinfo{journal}{Phys. Rev. E}} \textbf{\bibinfo{volume}{50}},
  \bibinfo{pages}{888--901} (\bibinfo{year}{1994}).
%\newblock %\urlprefix\url{https://link.aps.org/doi/10.1103/PhysRevE.50.888}.

\bibitem{Rigol2008Nature}
\bibinfo{author}{{Rigol}, M.}, \bibinfo{author}{{Dunjko}, V.} \&
  \bibinfo{author}{{Olshanii}, M.}
\newblock \bibinfo{title}{{Thermalization and its mechanism for generic
  isolated quantum systems}}.
\newblock \emph{\bibinfo{journal}{\nat}} \textbf{\bibinfo{volume}{452}},
  \bibinfo{pages}{854--858} (\bibinfo{year}{2008}).

\bibitem{DAlessio2016AP}
\bibinfo{author}{{D'Alessio}, L.}, \bibinfo{author}{{Kafri}, Y.},
  \bibinfo{author}{{Polkovnikov}, A.} \& \bibinfo{author}{{Rigol}, M.}
\newblock \bibinfo{title}{{From quantum chaos and eigenstate thermalization to
  statistical mechanics and thermodynamics}}.
\newblock \emph{\bibinfo{journal}{Advances in Physics}}
  \textbf{\bibinfo{volume}{65}}, \bibinfo{pages}{239--362}
  (\bibinfo{year}{2016}).

\bibitem{Anderson1958PR}
\bibinfo{author}{Anderson, P.~W.}
\newblock \bibinfo{title}{Absence of diffusion in certain random lattices}.
\newblock \emph{\bibinfo{journal}{Phys. Rev.}} \textbf{\bibinfo{volume}{109}},
  \bibinfo{pages}{1492--1505} (\bibinfo{year}{1958}).
%\newblock %\urlprefix\url{https://link.aps.org/doi/10.1103/PhysRev.109.1492}.

\bibitem{Basko2006AP}
\bibinfo{author}{Basko, D.}, \bibinfo{author}{Aleiner, I.} \&
  \bibinfo{author}{Altshuler, B.}
\newblock \bibinfo{title}{Metal–insulator transition in a weakly interacting
  many-electron system with localized single-particle states}.
\newblock \emph{\bibinfo{journal}{Annals of Physics}}
  \textbf{\bibinfo{volume}{321}}, \bibinfo{pages}{1126--1205}
  (\bibinfo{year}{2006}).
\newblock
  %\urlprefix\url{https://www.sciencedirect.com/science/article/pii/S0003491605002630}.

\bibitem{Rahul2015annu}
\bibinfo{author}{Nandkishore, R.} \& \bibinfo{author}{Huse, D.~A.}
\newblock \bibinfo{title}{Many-body localization and thermalization in quantum
  statistical mechanics}.
\newblock \emph{\bibinfo{journal}{Annual Review of Condensed Matter Physics}}
  \textbf{\bibinfo{volume}{6}}, \bibinfo{pages}{15--38} (\bibinfo{year}{2015}).
\newblock
  %\urlprefix\url{https://doi.org/10.1146/annurev-conmatphys-031214-014726}.

\bibitem{Abanin2019RMP}
\bibinfo{author}{Abanin, D.~A.}, \bibinfo{author}{Altman, E.},
  \bibinfo{author}{Bloch, I.} \& \bibinfo{author}{Serbyn, M.}
\newblock \bibinfo{title}{Colloquium: Many-body localization, thermalization,
  and entanglement}.
\newblock \emph{\bibinfo{journal}{Rev. Mod. Phys.}}
  \textbf{\bibinfo{volume}{91}}, \bibinfo{pages}{021001}
  (\bibinfo{year}{2019}).
\newblock
  %\urlprefix\url{https://link.aps.org/doi/10.1103/RevModPhys.91.021001}.

\bibitem{ALET2018CRP}
\bibinfo{author}{Alet, F.} \& \bibinfo{author}{Laflorencie, N.}
\newblock \bibinfo{title}{Many-body localization: {A}n introduction and
  selected topics}.
\newblock \emph{\bibinfo{journal}{Comptes Rendus Physique}}
  \textbf{\bibinfo{volume}{19}}, \bibinfo{pages}{498--525}
  (\bibinfo{year}{2018}).
\newblock
  %\urlprefix\url{https://www.sciencedirect.com/science/article/pii/S163107051830032X}.

\bibitem{Turner2018np}
\bibinfo{author}{{Turner}, C.~J.} \emph{et~al.}
\newblock \bibinfo{title}{{Weak ergodicity breaking from quantum many-body
  scars}}.
\newblock \emph{\bibinfo{journal}{Nature Physics}}
  \textbf{\bibinfo{volume}{14}}, \bibinfo{pages}{745--749}
  (\bibinfo{year}{2018}).

\bibitem{Serbyn2021np}
\bibinfo{author}{{Serbyn}, M.}, \bibinfo{author}{{Abanin}, D.~A.} \&
  \bibinfo{author}{{Papi{\'c}}, Z.}
\newblock \bibinfo{title}{{Quantum many-body scars and weak breaking of
  ergodicity}}.
\newblock \emph{\bibinfo{journal}{Nature Physics}}
  \textbf{\bibinfo{volume}{17}}, \bibinfo{pages}{675--685}
  (\bibinfo{year}{2021}).

\bibitem{Schreiber2015science}
\bibinfo{author}{Schreiber, M.} \emph{et~al.}
\newblock \bibinfo{title}{Observation of many-body localization of interacting
  fermions in a quasirandom optical lattice}.
\newblock \emph{\bibinfo{journal}{Science}} \textbf{\bibinfo{volume}{349}},
  \bibinfo{pages}{842--845} (\bibinfo{year}{2015}).
\newblock
  %\urlprefix\url{https://www.science.org/doi/abs/10.1126/science.aaa7432}.

\bibitem{Smith2016np}
\bibinfo{author}{{Smith}, J.} \emph{et~al.}
\newblock \bibinfo{title}{{Many-body localization in a quantum simulator with
  programmable random disorder}}.
\newblock \emph{\bibinfo{journal}{Nature Physics}}
  \textbf{\bibinfo{volume}{12}}, \bibinfo{pages}{907--911}
  (\bibinfo{year}{2016}).

\bibitem{Kai2018PRL}
\bibinfo{author}{Xu, K.} \emph{et~al.}
\newblock \bibinfo{title}{Emulating many-body localization with a
  superconducting quantum processor}.
\newblock \emph{\bibinfo{journal}{Phys. Rev. Lett.}}
  \textbf{\bibinfo{volume}{120}}, \bibinfo{pages}{050507}
  (\bibinfo{year}{2018}).
\newblock
  %\urlprefix\url{https://link.aps.org/doi/10.1103/PhysRevLett.120.050507}.

\bibitem{Roushan2017science}
\bibinfo{author}{Roushan, P.} \emph{et~al.}
\newblock \bibinfo{title}{Spectroscopic signatures of localization with
  interacting photons in superconducting qubits}.
\newblock \emph{\bibinfo{journal}{Science}} \textbf{\bibinfo{volume}{358}},
  \bibinfo{pages}{1175--1179} (\bibinfo{year}{2017}).
\newblock
  %\urlprefix\url{https://www.science.org/doi/abs/10.1126/science.aao1401}.

\bibitem{Xiao2021science}
\bibinfo{author}{Mi, X.} \emph{et~al.}
\newblock \bibinfo{title}{Information scrambling in computationally complex
  quantum circuits}.
\newblock \emph{\bibinfo{journal}{Science}} \textbf{\bibinfo{volume}{374}},
  \bibinfo{pages}{1479--1483} (\bibinfo{year}{2021}).
%\newblock %\urlprefix\url{https://doi.org/10.1126/science.abg5029}.

\bibitem{Neill2016np}
\bibinfo{author}{{Neill}, C.} \emph{et~al.}
\newblock \bibinfo{title}{{Ergodic dynamics and thermalization in an isolated
  quantum system}}.
\newblock \emph{\bibinfo{journal}{Nature Physics}}
  \textbf{\bibinfo{volume}{12}}, \bibinfo{pages}{1037--1041}
  (\bibinfo{year}{2016}).

\bibitem{Kaufman2016science}
\bibinfo{author}{{Kaufman}, A.~M.} \emph{et~al.}
\newblock \bibinfo{title}{{Quantum thermalization through entanglement in an
  isolated many-body system}}.
\newblock \emph{\bibinfo{journal}{Science}} \textbf{\bibinfo{volume}{353}},
  \bibinfo{pages}{794--800} (\bibinfo{year}{2016}).

\bibitem{Lukin2019science}
\bibinfo{author}{Lukin, A.} \emph{et~al.}
\newblock \bibinfo{title}{Probing entanglement in a many-body localized
  system}.
\newblock \emph{\bibinfo{journal}{Science}} \textbf{\bibinfo{volume}{364}},
  \bibinfo{pages}{256--260} (\bibinfo{year}{2019}).
\newblock
  %\urlprefix\url{https://www.science.org/doi/abs/10.1126/science.aau0818}.

\bibitem{Roeck2017PRB}
\bibinfo{author}{De~Roeck, W.} \& \bibinfo{author}{Huveneers, F. m.~c.}
\newblock \bibinfo{title}{Stability and instability towards delocalization in
  many-body localization systems}.
\newblock \emph{\bibinfo{journal}{Phys. Rev. B}} \textbf{\bibinfo{volume}{95}},
  \bibinfo{pages}{155129} (\bibinfo{year}{2017}).
%\newblock %\urlprefix\url{https://link.aps.org/doi/10.1103/PhysRevB.95.155129}.

\bibitem{Potirniche2019PRB}
\bibinfo{author}{Potirniche, I.-D.}, \bibinfo{author}{Banerjee, S.} \&
  \bibinfo{author}{Altman, E.}
\newblock \bibinfo{title}{Exploration of the stability of many-body
  localization in $d>1$}.
\newblock \emph{\bibinfo{journal}{Phys. Rev. B}} \textbf{\bibinfo{volume}{99}},
  \bibinfo{pages}{205149} (\bibinfo{year}{2019}).
%\newblock %\urlprefix\url{https://link.aps.org/doi/10.1103/PhysRevB.99.205149}.

\bibitem{Doggen2020PRL}
\bibinfo{author}{Doggen, E. V.~H.}, \bibinfo{author}{Gornyi, I.~V.},
  \bibinfo{author}{Mirlin, A.~D.} \& \bibinfo{author}{Polyakov, D.~G.}
\newblock \bibinfo{title}{Slow many-body delocalization beyond one dimension}.
\newblock \emph{\bibinfo{journal}{Phys. Rev. Lett.}}
  \textbf{\bibinfo{volume}{125}}, \bibinfo{pages}{155701}
  (\bibinfo{year}{2020}).
\newblock
  %\urlprefix\url{https://link.aps.org/doi/10.1103/PhysRevLett.125.155701}.

\bibitem{Potter2015PRX}
\bibinfo{author}{Potter, A.~C.}, \bibinfo{author}{Vasseur, R.} \&
  \bibinfo{author}{Parameswaran, S.~A.}
\newblock \bibinfo{title}{Universal properties of many-body delocalization
  transitions}.
\newblock \emph{\bibinfo{journal}{Phys. Rev. X}} \textbf{\bibinfo{volume}{5}},
  \bibinfo{pages}{031033} (\bibinfo{year}{2015}).
%\newblock %\urlprefix\url{https://link.aps.org/doi/10.1103/PhysRevX.5.031033}.

\bibitem{Khemani2017PRX}
\bibinfo{author}{Khemani, V.}, \bibinfo{author}{Lim, S.~P.},
  \bibinfo{author}{Sheng, D.~N.} \& \bibinfo{author}{Huse, D.~A.}
\newblock \bibinfo{title}{Critical properties of the many-body localization
  transition}.
\newblock \emph{\bibinfo{journal}{Phys. Rev. X}} \textbf{\bibinfo{volume}{7}},
  \bibinfo{pages}{021013} (\bibinfo{year}{2017}).
%\newblock %\urlprefix\url{https://link.aps.org/doi/10.1103/PhysRevX.7.021013}.

\bibitem{Dumitrescu2019PRB}
\bibinfo{author}{Dumitrescu, P.~T.}, \bibinfo{author}{Goremykina, A.},
  \bibinfo{author}{Parameswaran, S.~A.}, \bibinfo{author}{Serbyn, M.} \&
  \bibinfo{author}{Vasseur, R.}
\newblock \bibinfo{title}{Kosterlitz-{T}houless scaling at many-body
  localization phase transitions}.
\newblock \emph{\bibinfo{journal}{Phys. Rev. B}} \textbf{\bibinfo{volume}{99}},
  \bibinfo{pages}{094205} (\bibinfo{year}{2019}).
%\newblock %\urlprefix\url{https://link.aps.org/doi/10.1103/PhysRevB.99.094205}.

\bibitem{Choi2016science}
\bibinfo{author}{yoon Choi, J.} \emph{et~al.}
\newblock \bibinfo{title}{Exploring the many-body localization transition in
  two dimensions}.
\newblock \emph{\bibinfo{journal}{Science}} \textbf{\bibinfo{volume}{352}},
  \bibinfo{pages}{1547--1552} (\bibinfo{year}{2016}).
\newblock
  %\urlprefix\url{https://www.science.org/doi/abs/10.1126/science.aaf8834}.
%\newblock \eprint{https://www.science.org/doi/pdf/10.1126/science.aaf8834}.

\bibitem{BordiaPRX2017}
\bibinfo{author}{Bordia, P.} \emph{et~al.}
\newblock \bibinfo{title}{Probing slow relaxation and many-body localization in
  two-dimensional quasiperiodic systems}.
\newblock \emph{\bibinfo{journal}{Phys. Rev. X}} \textbf{\bibinfo{volume}{7}},
  \bibinfo{pages}{041047} (\bibinfo{year}{2017}).
%\newblock %\urlprefix\url{https://link.aps.org/doi/10.1103/PhysRevX.7.041047}.

\bibitem{Rispoli2019science}
\bibinfo{author}{{Rispoli}, M.} \emph{et~al.}
\newblock \bibinfo{title}{{Quantum critical behaviour at the many-body
  localization transition}}.
\newblock \emph{\bibinfo{journal}{\nat}} \textbf{\bibinfo{volume}{573}},
  \bibinfo{pages}{385--389} (\bibinfo{year}{2019}).

\bibitem{Welsh2018JoConMatt}
\bibinfo{author}{Welsh, S.} \& \bibinfo{author}{Logan, D.~E.}
\newblock \bibinfo{title}{Simple probability distributions on a {F}ock-space
  lattice}.
\newblock \emph{\bibinfo{journal}{Journal of Physics: Condensed Matter}}
  \textbf{\bibinfo{volume}{30}}, \bibinfo{pages}{405601}
  (\bibinfo{year}{2018}).
%\newblock %\urlprefix\url{https://doi.org/10.1088/1361-648x/aadd35}.

\bibitem{Altshuler1997PRL}
\bibinfo{author}{Altshuler, B.~L.}, \bibinfo{author}{Gefen, Y.},
  \bibinfo{author}{Kamenev, A.} \& \bibinfo{author}{Levitov, L.~S.}
\newblock \bibinfo{title}{Quasiparticle lifetime in a finite system: A
  nonperturbative approach}.
\newblock \emph{\bibinfo{journal}{Phys. Rev. Lett.}}
  \textbf{\bibinfo{volume}{78}}, \bibinfo{pages}{2803--2806}
  (\bibinfo{year}{1997}).
%\newblock %\urlprefix\url{https://link.aps.org/doi/10.1103/PhysRevLett.78.2803}.

\bibitem{Nicolas2019PRL}
\bibinfo{author}{Mac\'e, N.}, \bibinfo{author}{Alet, F.} \&
  \bibinfo{author}{Laflorencie, N.}
\newblock \bibinfo{title}{Multifractal scalings across the many-body
  localization transition}.
\newblock \emph{\bibinfo{journal}{Phys. Rev. Lett.}}
  \textbf{\bibinfo{volume}{123}}, \bibinfo{pages}{180601}
  (\bibinfo{year}{2019}).
\newblock
  %\urlprefix\url{https://link.aps.org/doi/10.1103/PhysRevLett.123.180601}.

\bibitem{Logan2019}
\bibinfo{author}{Logan, D.~E.} \& \bibinfo{author}{Welsh, S.}
\newblock \bibinfo{title}{Many-body localization in {F}ock space: A local
  perspective}.
\newblock \emph{\bibinfo{journal}{Phys. Rev. B}} \textbf{\bibinfo{volume}{99}},
  \bibinfo{pages}{045131} (\bibinfo{year}{2019}).
%\newblock %\urlprefix\url{https://link.aps.org/doi/10.1103/PhysRevB.99.045131}.

\bibitem{Tomasi2019PRB}
\bibinfo{author}{De~Tomasi, G.}, \bibinfo{author}{Hetterich, D.},
  \bibinfo{author}{Sala, P.} \& \bibinfo{author}{Pollmann, F.}
\newblock \bibinfo{title}{Dynamics of strongly interacting systems: From
  {F}ock-space fragmentation to many-body localization}.
\newblock \emph{\bibinfo{journal}{Phys. Rev. B}}
  \textbf{\bibinfo{volume}{100}}, \bibinfo{pages}{214313}
  (\bibinfo{year}{2019}).
%\newblock %\urlprefix\url{https://link.aps.org/doi/10.1103/PhysRevB.100.214313}.

\bibitem{Luca2014PRL}
\bibinfo{author}{De~Luca, A.}, \bibinfo{author}{Altshuler, B.~L.},
  \bibinfo{author}{Kravtsov, V.~E.} \& \bibinfo{author}{Scardicchio, A.}
\newblock \bibinfo{title}{Anderson localization on the {B}ethe lattice:
  {N}onergodicity of extended states}.
\newblock \emph{\bibinfo{journal}{Phys. Rev. Lett.}}
  \textbf{\bibinfo{volume}{113}}, \bibinfo{pages}{046806}
  (\bibinfo{year}{2014}).
\newblock
  %\urlprefix\url{https://link.aps.org/doi/10.1103/PhysRevLett.113.046806}.

\bibitem{Tomasi2019spp}
\bibinfo{author}{Tomasi, G.~D.}, \bibinfo{author}{Amini, M.},
  \bibinfo{author}{Bera, S.}, \bibinfo{author}{Khaymovich, I.~M.} \&
  \bibinfo{author}{Kravtsov, V.~E.}
\newblock \bibinfo{title}{{Survival probability in Generalized
  {R}osenzweig-{P}orter random matrix ensemble}}.
\newblock \emph{\bibinfo{journal}{SciPost Phys.}} \textbf{\bibinfo{volume}{6}},
  \bibinfo{pages}{014} (\bibinfo{year}{2019}).
%\newblock %\urlprefix\url{https://scipost.org/10.21468/SciPostPhys.6.1.014}.

\bibitem{Wang2021PRL}
\bibinfo{author}{Wang, Y.}, \bibinfo{author}{Cheng, C.}, \bibinfo{author}{Liu,
  X.-J.} \& \bibinfo{author}{Yu, D.}
\newblock \bibinfo{title}{Many-body critical phase: {E}xtended and nonthermal}.
\newblock \emph{\bibinfo{journal}{Phys. Rev. Lett.}}
  \textbf{\bibinfo{volume}{126}}, \bibinfo{pages}{080602}
  (\bibinfo{year}{2021}).
\newblock
  %\urlprefix\url{https://link.aps.org/doi/10.1103/PhysRevLett.126.080602}.

\bibitem{Benalcazar2017science}
\bibinfo{author}{Benalcazar, W.~A.}, \bibinfo{author}{Bernevig, B.~A.} \&
  \bibinfo{author}{Hughes, T.~L.}
\newblock \bibinfo{title}{Quantized electric multipole insulators}.
\newblock \emph{\bibinfo{journal}{Science}} \textbf{\bibinfo{volume}{357}},
  \bibinfo{pages}{61--66} (\bibinfo{year}{2017}).
\newblock
  %\urlprefix\url{https://www.science.org/doi/abs/10.1126/science.aah6442}.

\bibitem{Karamlou2022npj}
\bibinfo{author}{{Karamlou}, A.~H.} \emph{et~al.}
\newblock \bibinfo{title}{{Quantum transport and localization in 1d and 2d
  tight-binding lattices}}.
\newblock \emph{\bibinfo{journal}{npj Quantum Information}}
  \textbf{\bibinfo{volume}{8}}, \bibinfo{pages}{35} (\bibinfo{year}{2022}).

\bibitem{Tomasi2021PRB}
\bibinfo{author}{De~Tomasi, G.}, \bibinfo{author}{Khaymovich, I.~M.},
  \bibinfo{author}{Pollmann, F.} \& \bibinfo{author}{Warzel, S.}
\newblock \bibinfo{title}{Rare thermal bubbles at the many-body localization
  transition from the {F}ock space point of view}.
\newblock \emph{\bibinfo{journal}{Phys. Rev. B}}
  \textbf{\bibinfo{volume}{104}}, \bibinfo{pages}{024202}
  (\bibinfo{year}{2021}).
%\newblock %\urlprefix\url{https://link.aps.org/doi/10.1103/PhysRevB.104.024202}.

\bibitem{zhang2022ArXiv}
\bibinfo{author}{{Zhang}, P.} \emph{et~al.}
\newblock \bibinfo{title}{Many-body hilbert space scarring on a superconducting
  processor}.
\newblock \emph{\bibinfo{journal}{Nature Physics}} \bibinfo{pages}{1--6}
  (\bibinfo{year}{2022}). https://doi.org/10.1038/s41567-022-01784-9

\bibitem{Yan2019Science}
\bibinfo{author}{Yan, Z.} \emph{et~al.}
\newblock \bibinfo{title}{Strongly correlated quantum walks with a 12-qubit
  superconducting processor}.
\newblock \emph{\bibinfo{journal}{Science}} \textbf{\bibinfo{volume}{364}},
  \bibinfo{pages}{753--756} (\bibinfo{year}{2019}).
\newblock
  %\urlprefix\url{https://www.science.org/doi/abs/10.1126/science.aaw1611}.

\bibitem{Braumuller2021NP}
\bibinfo{author}{{Braum{\"u}ller}, J.} \emph{et~al.}
\newblock \bibinfo{title}{{Probing quantum information propagation with
  out-of-time-ordered correlators}}.
\newblock \emph{\bibinfo{journal}{Nature Physics}}
  \textbf{\bibinfo{volume}{18}}, \bibinfo{pages}{172--178}
  (\bibinfo{year}{2021}).

\bibitem{Aoki1980JPC}
\bibinfo{author}{Aoki, H.}
\newblock \bibinfo{title}{Real-space renormalisation-group theory for anderson
  localisation: decimation method for electron systems}.
\newblock \emph{\bibinfo{journal}{Journal of Physics C: Solid State Physics}}
  \textbf{\bibinfo{volume}{13}}, \bibinfo{pages}{3369--3386}
  (\bibinfo{year}{1980}).
%\newblock %\urlprefix\url{https://doi.org/10.1088/0022-3719/13/18/006}.

\bibitem{Pietracaprina2021AP}
\bibinfo{author}{Pietracaprina, F.} \& \bibinfo{author}{Laflorencie, N.}
\newblock \bibinfo{title}{Hilbert-space fragmentation, multifractality, and
  many-body localization}.
\newblock \emph{\bibinfo{journal}{Annals of Physics}}
  \textbf{\bibinfo{volume}{435}}, \bibinfo{pages}{168502}
  (\bibinfo{year}{2021}).
\newblock
  %\urlprefix\url{https://www.sciencedirect.com/science/article/pii/S0003491621001081}.
\newblock \bibinfo{note}{Special Issue on Localisation 2020}.

\bibitem{Suntajs2020PRE}
\bibinfo{author}{\ifmmode~\check{S}\else \v{S}\fi{}untajs, J.},
  \bibinfo{author}{Bon\ifmmode~\check{c}\else \v{c}\fi{}a, J.},
  \bibinfo{author}{Prosen, T. c.~v.} \& \bibinfo{author}{Vidmar, L.}
\newblock \bibinfo{title}{Quantum chaos challenges many-body localization}.
\newblock \emph{\bibinfo{journal}{Phys. Rev. E}}
  \textbf{\bibinfo{volume}{102}}, \bibinfo{pages}{062144}
  (\bibinfo{year}{2020}).
%\newblock %\urlprefix\url{https://link.aps.org/doi/10.1103/PhysRevE.102.062144}.

\bibitem{Jonas2014PRL}
\bibinfo{author}{Kj\"all, J.~A.}, \bibinfo{author}{Bardarson, J.~H.} \&
  \bibinfo{author}{Pollmann, F.}
\newblock \bibinfo{title}{Many-body localization in a disordered quantum
  {I}sing chain}.
\newblock \emph{\bibinfo{journal}{Phys. Rev. Lett.}}
  \textbf{\bibinfo{volume}{113}}, \bibinfo{pages}{107204}
  (\bibinfo{year}{2014}).
\newblock
  %\urlprefix\url{https://link.aps.org/doi/10.1103/PhysRevLett.113.107204}.

\bibitem{Luitz2015PRB}
\bibinfo{author}{Luitz, D.~J.}, \bibinfo{author}{Laflorencie, N.} \&
  \bibinfo{author}{Alet, F.}
\newblock \bibinfo{title}{Many-body localization edge in the random-field
  {H}eisenberg chain}.
\newblock \emph{\bibinfo{journal}{Phys. Rev. B}} \textbf{\bibinfo{volume}{91}},
  \bibinfo{pages}{081103} (\bibinfo{year}{2015}).
%\newblock %\urlprefix\url{https://link.aps.org/doi/10.1103/PhysRevB.91.081103}.

\bibitem{Evers2008RMP}
\bibinfo{author}{Evers, F.} \& \bibinfo{author}{Mirlin, A.~D.}
\newblock \bibinfo{title}{Anderson transitions}.
\newblock \emph{\bibinfo{journal}{Rev. Mod. Phys.}}
  \textbf{\bibinfo{volume}{80}}, \bibinfo{pages}{1355--1417}
  (\bibinfo{year}{2008}).
%\newblock %\urlprefix\url{https://link.aps.org/doi/10.1103/RevModPhys.80.1355}.

\bibitem{Luitz2014PRB}
\bibinfo{author}{Luitz, D.~J.}, \bibinfo{author}{Alet, F.} \&
  \bibinfo{author}{Laflorencie, N.}
\newblock \bibinfo{title}{Universal behavior beyond multifractality in quantum
  many-body systems}.
\newblock \emph{\bibinfo{journal}{Phys. Rev. Lett.}}
  \textbf{\bibinfo{volume}{112}}, \bibinfo{pages}{057203}
  (\bibinfo{year}{2014}).
\newblock
  %\urlprefix\url{https://link.aps.org/doi/10.1103/PhysRevLett.112.057203}.

\bibitem{Suntajs2020PRB}
\bibinfo{author}{\ifmmode~\check{S}\else \v{S}\fi{}untajs, J.},
  \bibinfo{author}{Bon\ifmmode~\check{c}\else \v{c}\fi{}a, J.},
  \bibinfo{author}{Prosen, T. c.~v.} \& \bibinfo{author}{Vidmar, L.}
\newblock \bibinfo{title}{Ergodicity breaking transition in finite disordered
  spin chains}.
\newblock \emph{\bibinfo{journal}{Phys. Rev. B}}
  \textbf{\bibinfo{volume}{102}}, \bibinfo{pages}{064207}
  (\bibinfo{year}{2020}).
%\newblock %\urlprefix\url{https://link.aps.org/doi/10.1103/PhysRevB.102.064207}.

\bibitem{Hauke2015PRB}
\bibinfo{author}{Hauke, P.} \& \bibinfo{author}{Heyl, M.}
\newblock \bibinfo{title}{Many-body localization and quantum ergodicity in
  disordered long-range {I}sing models}.
\newblock \emph{\bibinfo{journal}{Phys. Rev. B}} \textbf{\bibinfo{volume}{92}},
  \bibinfo{pages}{134204} (\bibinfo{year}{2015}).
%\newblock %\urlprefix\url{https://link.aps.org/doi/10.1103/PhysRevB.92.134204}.

\bibitem{Guo2021PRL}
\bibinfo{author}{Guo, Q.} \emph{et~al.}
\newblock \bibinfo{title}{Stark many-body localization on a superconducting
  quantum processor}.
\newblock \emph{\bibinfo{journal}{Phys. Rev. Lett.}}
  \textbf{\bibinfo{volume}{127}}, \bibinfo{pages}{240502}
  (\bibinfo{year}{2021}).
\newblock
  %\urlprefix\url{https://link.aps.org/doi/10.1103/PhysRevLett.127.240502}.

\bibitem{Page1993}
\bibinfo{author}{Page, D.~N.}
\newblock \bibinfo{title}{Average entropy of a subsystem}.
\newblock \emph{\bibinfo{journal}{Phys. Rev. Lett.}}
  \textbf{\bibinfo{volume}{71}}, \bibinfo{pages}{1291--1294}
  (\bibinfo{year}{1993}).
%\newblock %\urlprefix\url{https://link.aps.org/doi/10.1103/PhysRevLett.71.1291}.

\bibitem{Goremykina2019PRL}
\bibinfo{author}{Goremykina, A.}, \bibinfo{author}{Vasseur, R.} \&
  \bibinfo{author}{Serbyn, M.}
\newblock \bibinfo{title}{Analytically solvable renormalization group for the
  many-body localization transition}.
\newblock \emph{\bibinfo{journal}{Phys. Rev. Lett.}}
  \textbf{\bibinfo{volume}{122}}, \bibinfo{pages}{040601}
  (\bibinfo{year}{2019}).
\newblock
  %\urlprefix\url{https://link.aps.org/doi/10.1103/PhysRevLett.122.040601}.

\bibitem{Foo2022Arxiv}
\bibinfo{author}{{Foo}, D.~C.~W.}, \bibinfo{author}{{Swain}, N.},
  \bibinfo{author}{{Sengupta}, P.}, \bibinfo{author}{{Lemari{\'e}}, G.} \&
  \bibinfo{author}{{Adam}, S.}
\newblock \bibinfo{title}{{A stabilization mechanism for many-body localization
  in two dimensions}}.
\newblock \emph{\bibinfo{journal}{arXiv:2202.09072}}  (\bibinfo{year}{2022}).
%\newblock %\urlprefix\url{https://arxiv.org/abs/2202.09072}.

\bibitem{Wahl2018NP}
\bibinfo{author}{Wahl, T.~B.}, \bibinfo{author}{Pal, A.} \&
  \bibinfo{author}{Simon, S.~H.}
\newblock \bibinfo{title}{Signatures of the many-body localized regime in two
  dimensions}.
\newblock \emph{\bibinfo{journal}{Nature Physics}}
  \textbf{\bibinfo{volume}{15}}, \bibinfo{pages}{164--169}
  (\bibinfo{year}{2018}).
%\newblock %\urlprefix\url{https://doi.org/10.1038%2Fs41567-018-0339-x}.

\bibitem{Kshetrimayum2020}
\bibinfo{author}{Kshetrimayum, A.}, \bibinfo{author}{Goihl, M.} \&
  \bibinfo{author}{Eisert, J.}
\newblock \bibinfo{title}{Time evolution of many-body localized systems in two
  spatial dimensions}.
\newblock \emph{\bibinfo{journal}{Phys. Rev. B}}
  \textbf{\bibinfo{volume}{102}}, \bibinfo{pages}{235132}
  (\bibinfo{year}{2020}).
%\newblock %\urlprefix\url{https://link.aps.org/doi/10.1103/PhysRevB.102.235132}.

\bibitem{eveniaut2020PRR}
\bibinfo{author}{Th\'eveniaut, H.}, \bibinfo{author}{Lan, Z.},
  \bibinfo{author}{Meyer, G.} \& \bibinfo{author}{Alet, F.}
\newblock \bibinfo{title}{Transition to a many-body localized regime in a
  two-dimensional disordered quantum dimer model}.
\newblock \emph{\bibinfo{journal}{Phys. Rev. Research}}
  \textbf{\bibinfo{volume}{2}}, \bibinfo{pages}{033154} (\bibinfo{year}{2020}).
\newblock
  %\urlprefix\url{https://link.aps.org/doi/10.1103/PhysRevResearch.2.033154}.

\bibitem{Chertkov2021PRL}
\bibinfo{author}{Chertkov, E.}, \bibinfo{author}{Villalonga, B.} \&
  \bibinfo{author}{Clark, B.~K.}
\newblock \bibinfo{title}{Numerical evidence for many-body localization in two
  and three dimensions}.
\newblock \emph{\bibinfo{journal}{Phys. Rev. Lett.}}
  \textbf{\bibinfo{volume}{126}}, \bibinfo{pages}{180602}
  (\bibinfo{year}{2021}).
\newblock
  %\urlprefix\url{https://link.aps.org/doi/10.1103/PhysRevLett.126.180602}.

\bibitem{Luitz2016PRB}
\bibinfo{author}{Luitz, D.~J.}, \bibinfo{author}{Laflorencie, N.} \&
  \bibinfo{author}{Alet, F.}
\newblock \bibinfo{title}{Extended slow dynamical regime close to the many-body
  localization transition}.
\newblock \emph{\bibinfo{journal}{Phys. Rev. B}} \textbf{\bibinfo{volume}{93}},
  \bibinfo{pages}{060201} (\bibinfo{year}{2016}).
%\newblock %\urlprefix\url{https://link.aps.org/doi/10.1103/PhysRevB.93.060201}.

\bibitem{Yan2018PRAp}
\bibinfo{author}{Yan, F.} \emph{et~al.}
\newblock \bibinfo{title}{Tunable coupling scheme for implementing
  high-fidelity two-qubit gates}.
\newblock \emph{\bibinfo{journal}{Phys. Rev. Applied}}
  \textbf{\bibinfo{volume}{10}}, \bibinfo{pages}{054062}
  (\bibinfo{year}{2018}).
\newblock
  %\urlprefix\url{https://link.aps.org/doi/10.1103/PhysRevApplied.10.054062}.

\bibitem{zhang2022Nature}
\bibinfo{author}{Zhang, X.} \emph{et~al.}
\newblock \bibinfo{title}{Digital quantum simulation of {F}loquet
  symmetry-protected topological phases}.
\newblock \emph{\bibinfo{journal}{Nature}} \textbf{\bibinfo{volume}{607}},
  \bibinfo{pages}{468--473} (\bibinfo{year}{2022}).
%\newblock %\urlprefix\url{https://www.nature.com/articles/s41586-022-04854-3}.

\bibitem{Bravyi_2011}
\bibinfo{author}{Bravyi, S.}, \bibinfo{author}{DiVincenzo, D.~P.} \&
  \bibinfo{author}{Loss, D.}
\newblock \bibinfo{title}{Schrieffer{\textendash}{W}olff transformation for
  quantum many-body systems}.
\newblock \emph{\bibinfo{journal}{Annals of Physics}}
  \textbf{\bibinfo{volume}{326}}, \bibinfo{pages}{2793--2826}
  (\bibinfo{year}{2011}).
%\newblock %\urlprefix\url{https://doi.org/10.1016%2Fj.aop.2011.06.004}.

\bibitem{Guo2021np}
\bibinfo{author}{{Guo}, Q.} \emph{et~al.}
\newblock \bibinfo{title}{{Observation of energy-resolved many-body
  localization}}.
\newblock \emph{\bibinfo{journal}{Nature Physics}}
  \textbf{\bibinfo{volume}{17}}, \bibinfo{pages}{234--239}
  (\bibinfo{year}{2021}).

\bibitem{articlescar}
\bibinfo{author}{Turner, C.}, \bibinfo{author}{Michailidis, A.},
  \bibinfo{author}{Abanin, D.}, \bibinfo{author}{Serbyn, M.} \&
  \bibinfo{author}{Papic, Z.}
\newblock \bibinfo{title}{Weak ergodicity breaking from quantum many-body
  scars}.
\newblock \emph{\bibinfo{journal}{Nature Physics}}
  \textbf{\bibinfo{volume}{14}} (\bibinfo{year}{2018}).

\bibitem{Hamazaki_2018}
\bibinfo{author}{Hamazaki, R.} \& \bibinfo{author}{Ueda, M.}
\newblock \bibinfo{title}{Atypicality of most few-body observables}.
\newblock \emph{\bibinfo{journal}{Physical Review Letters}}
  \textbf{\bibinfo{volume}{120}} (\bibinfo{year}{2018}).
%\newblock %\urlprefix\url{https://doi.org/10.1103%2Fphysrevlett.120.080603}.

\bibitem{GUHR1998189}
\bibinfo{author}{Guhr, T.}, \bibinfo{author}{Müller–Groeling, A.} \&
  \bibinfo{author}{Weidenmüller, H.~A.}
\newblock \bibinfo{title}{Random-matrix theories in quantum physics: common
  concepts}.
\newblock \emph{\bibinfo{journal}{Physics Reports}}
  \textbf{\bibinfo{volume}{299}}, \bibinfo{pages}{189--425}
  (\bibinfo{year}{1998}).
\newblock
  %\urlprefix\url{https://www.sciencedirect.com/science/article/pii/S0370157397000884}.

\bibitem{IZRAILEV1990299}
\bibinfo{author}{Izrailev, F.~M.}
\newblock \bibinfo{title}{Simple models of quantum chaos: Spectrum and
  eigenfunctions}.
\newblock \emph{\bibinfo{journal}{Physics Reports}}
  \textbf{\bibinfo{volume}{196}}, \bibinfo{pages}{299--392}
  (\bibinfo{year}{1990}).
\newblock
  %\urlprefix\url{https://www.sciencedirect.com/science/article/pii/037015739090067C}.

\bibitem{Evers_2008}
\bibinfo{author}{Evers, F.} \& \bibinfo{author}{Mirlin, A.~D.}
\newblock \bibinfo{title}{Anderson transitions}.
\newblock \emph{\bibinfo{journal}{Reviews of Modern Physics}}
  \textbf{\bibinfo{volume}{80}}, \bibinfo{pages}{1355--1417}
  (\bibinfo{year}{2008}).
%\newblock %\urlprefix\url{https://doi.org/10.1103%2Frevmodphys.80.1355}.

\bibitem{Google2020Arxiv}
\bibinfo{author}{Quantum, G.~A.} \& \bibinfo{author}{collaborators}.
\newblock \bibinfo{title}{{Observation of separated dynamics of charge and spin
  in the {F}ermi-{H}ubbard model}}.
\newblock \emph{\bibinfo{journal}{arXiv:2010.07965}}  (\bibinfo{year}{2020}).
%\newblock %\urlprefix\url{https://arxiv.org/abs/2010.07965}.

\end{thebibliography}

\end{document}